\documentclass[aps,prb,reprint,groupedaddress,amsmath,amssymb,superscriptaddress]{revtex4-2}
\usepackage{graphicx}
\usepackage{float}
\usepackage{multirow}
\usepackage{natbib,twoopt}
\usepackage[breaklinks=true]{hyperref}
\usepackage{xcolor}
\usepackage{amssymb}
\usepackage{stackengine,graphicx}
\usepackage{amsmath}
\usepackage{hyperref}
\usepackage{color}
\usepackage{stackengine}
\usepackage{subfigure}
\usepackage{verbatim}
\usepackage{bbold}
\usepackage{stmaryrd} 
\usepackage{overpic}
\usepackage[normalem]{ulem}
\usepackage{soul}

\newcommand{\ds}[2]{\frac{\partial^2 #1}{\partial #2^2}}
\newcommand{\eps}{\varepsilon}

\newcommand{\bas}{{\bf e}_z}
\newcommand{\om}{\omega}

\newcommand{\cd}{\tilde\chi}
\newcommand{\vc}[1]{\boldsymbol{\mathrm{#1}}}


\begin{document}

\title{Emergence of dual axion response in condensed matter}

\author{Elina Kokurina}
\affiliation{School of Physics and Engineering, ITMO University, Saint  Petersburg 197101, Russia}
\author{Dmitry Vagin}
\affiliation{School of Physics and Engineering, ITMO University, Saint  Petersburg 197101, Russia}
\author{Eduardo Barredo-Alamilla}
\affiliation{School of Physics and Engineering, ITMO University, Saint  Petersburg 197101, Russia}

\author{Maxim A. Gorlach}
\email{m.gorlach@metalab.ifmo.ru}
\affiliation{School of Physics and Engineering, ITMO University, Saint  Petersburg 197101, Russia}

\begin{abstract}
Recently, it was predicted that nonreciprocal magneto-electric effect in antiferromagnetic multilayered metamaterials occurs in two distinct versions. One is the conventional axion response, while another one is {\it dual axion response} captured by electrodynamics with magnetic charge. Here we investigate a model condensed matter system  of spins coupled through antiferromagnetic exchange interaction and derive its effective electromagnetic properties. We predict that this system gives rise to the emergent dual axion field, support our conclusion by numerical simulations and put forward  candidate materials.
\end{abstract}

\maketitle

\section{Introduction}

Nonreciprocal electromagnetic phenomena open interesting prospects in light manipulation by rendering the transmission coefficients in forward and backward direction unequal~\cite{Landau,Caloz2018,Asadchy2020}. A celebrated example of such kind is the Faraday effect in the medium with nonzero magnetization. Yet, the nonreciprocity persists even if the net magnetization is compensated, provided the structure includes several magnetic sublattices shifted relative to each other. An important example of such kind is antiferromagnetic materials. While known for a long time, they have recently appeared in the spotlight after the prediction of magnetic materials with compensated magnetization and spin-split bands~--  altermagnets~\cite{Smejkal2022,Smejkal2022a,Jungwirth2026}. 

Some of antiferromagnetic materials give rise to the non-reciprocal magneto-electric effect~\cite{Dzyaloshinskii1960,Astrov1960,Krichevtsov1993,Kimel2025} also known as an {\it axion response}~\cite{Nenno2020,Sekine2021}. Electrodynamics of such media is captured by the material equations of the form~\cite{Wilczek87,Nenno2020}
\begin{gather}
    {\bf D}=\varepsilon\,{\bf E}+\chi\,{\bf B}\:,\label{eq:AxionMat1}\\
    {\bf H}=-\chi\,{\bf E}+\mu^{-1}\,{\bf B}\:,\label{eq:AxionMat2}
\end{gather}
where $\eps$ and $\mu$ are the usual permittivity and permeability, while real coefficient $\chi$ quantifies the emergent axion response, being odd under spatial inversion and odd under time reversal. Interestingly, Maxwell's equations in such media written in terms of ${\bf E}$ and ${\bf B}$ fields coincide with the equations of axion electrodynamics capturing the physics of hypothetical axions~\cite{Wilczek87,Nenno2020,Sekine2021}.

Another remarkable property of Eqs.~\eqref{eq:AxionMat1}, \eqref{eq:AxionMat2} is that constant $\chi$ is not manifested in the bulk of the medium and only leads to the modification of the boundary conditions. Despite that, $\chi$ can be viewed as a bulk property, because once $\chi$ becomes coordinate-dependent or tensorial, it affects wave propagation in the bulk. Such behavior of $\chi$ makes it a very special type of linear response.


Furthermore, in strong 3D topological insulators such as Bi$_2$Te$_3$ or Bi$_2$Se$_3$ the axion response $\chi$ takes quantized values in units of the fine structure constant $\alpha$, playing the role of a topological invariant~\cite{Essin2009,Vazifeh2010,Armitage2016}. A related area are magnetic topological insulators, where nontrivial magnetization induces the topological order~\cite{Tokura2019,Bernevig2022,Li2023}.

Such physics has motivated much interest to the axion electrodynamics in solids~\cite{Nenno2020,SuYangXu2023,SuYangXu}. However, the available axion responses $\chi$ are typically quite small, below $10^{-3}\div 10^{-2}$, limiting the strength of available physical phenomena. Few years ago, that limitation has been challenged by the prediction~\cite{Shaposhnikov2023} that artificially structured media can feature much stronger axion responses $\chi\sim 1$. This theoretical conjecture~\cite{Shaposhnikov2023} was supported both by numerical simulations~\cite{Asadchy2024} and recent experiments~\cite{ShuangZhang2025,Asadchy2025} opening prospects to harness strong axion response for nonreciprocal photonics. 

In the context of photonics, the materials with such response are widely known as Tellegen media. While individual Tellegen meta-atoms were suggested long ago~\cite{Tretyakov2003}, clear examples of Tellegen response in composite photonic structures were lacking until recently~\cite{Shaposhnikov2023}, which stimulated renewed theoretical~\cite{Filipini2025,ShanhuiFan}, numerical~\cite{Asadchy2024} and, eventually, experimental~\cite{ShuangZhang2025,Asadchy2025,Liu2025,Lai2025} interest.

A very recent twist in the field is the prediction that nonreciprocal magneto-electric effect comes in two distinct versions. One is the conventional axion electrodynamics, while another is its dual-symmetric counterpart captured by the equations~\cite{Seidov2025}:
\begin{gather}
\nabla\times\vc{H}=\frac{1}{c}\frac{\partial {\bf D}}{\partial t} +\frac{4\pi}{c}\vc{j}\:,\label{eq:Maxwell1}\\
\nabla\cdot \vc{D}=4\pi\rho\:,\label{eq:Maxwell2}\\
\nabla\times(\vc{E}+\tilde{\chi}\vc{H})=-\frac{1}{c}\frac{\partial}{\partial t}(\vc{B}-\tilde{\chi}\vc{D})\:,\label{eq:Maxwell3}\\
\nabla\cdot (\vc{B}-\tilde{\chi}\vc{D})=0\:,\label{eq:Maxwell4}
\end{gather}
where $\cd$ is termed {\it dual axion response}. In this effective description, the fields ${\bf E}$, ${\bf D}$, ${\bf H}$ and ${\bf B}$ are understood as the averaged fields inside the structure related to each other via ${\bf D}=\varepsilon\,{\bf E}$ and ${\bf B}=\mu\,{\bf H}$. 

Similarly to the axion case, $\cd$ manifests itself only at the boundary of the medium, breaks time-reversal and spatial inversion symmetries and couples to the incident plane waves similarly to the axion electrodynamics. However, the interaction with the external sources introduced inside the medium is different from the axion case and provides a tool to distinguish the two versions of nonreciprocal magneto-electric effect~\cite{Seidov2025}. Counter-intuitively, this physics [Eqs.~\eqref{eq:Maxwell1}-\eqref{eq:Maxwell4}] is captured by electrodynamics with magnetic charge which is uncommon in condensed matter and photonics.

\begin{figure}[t]
    \centering
    \includegraphics[width=0.7\linewidth]{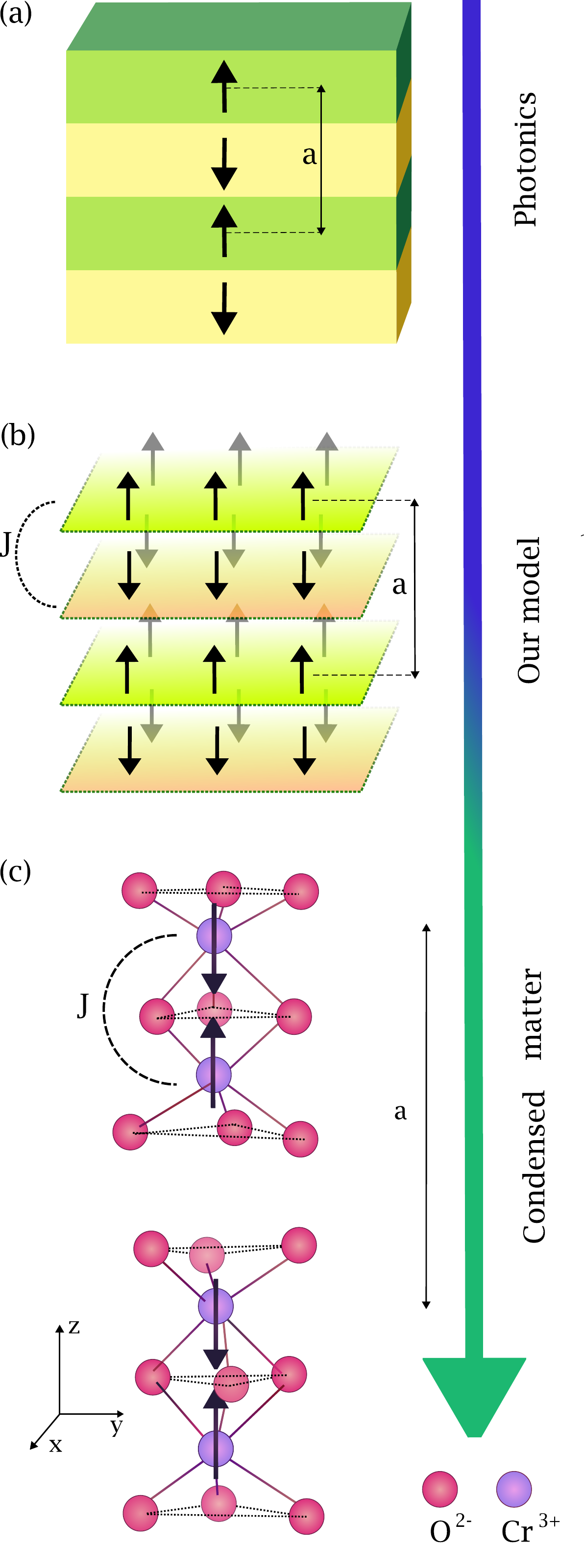}
    \caption{Hierarchy of the models featuring nonreciprocal magneto-electric effect. (a) Multilayered metamaterial composed of gyrotropic layers with out-of-plane magnetization features axion~\cite{Shaposhnikov2023} or dual axion~\cite{Seidov2025} response depending on the properties of the layers. (b) Our simplified model of the spin lattice with antiferromagnetic exchange coupling between the spins in the neighboring layers. (c) Realistic condensed matter structure Cr$_2$O$_3$ featuring an antiferromagnetic arrangement of Cr spins~\cite{Shiratsuchi} and promising for observing dual axion response.} 
    \label{fig:spin chain}
\end{figure}

This exotic response has been predicted~\cite{Seidov2025} for a metamaterial structure composed of gyrotropic layers with out-of-plane magnetization [Fig.~\ref{fig:spin chain}(a)]. On the other hand, more complicated condensed matter systems such as Cr$_2$O$_3$ or MnBi$_2$Te$_4$ feature a similar antiferromagnetic pattern of magnetization~\cite{Nenno2020}. By now, it remains unclear whether newly predicted dual axion response can occur in condensed matter and what are the candidate materials. 

In this Article, we fill this gap studying a model system of spins interacting with each other via Heisenberg-type antiferromagnetic exchange coupling [Fig.~\ref{fig:spin chain}(b)] which bridges the gap between the idealized metamaterials investigated previously [Fig.~\ref{fig:spin chain}(a)] and condensed matter structures [Fig.~\ref{fig:spin chain}(c)]. On one hand, our system resembles a simpler classical multilayered metamaterial and is expected to feature a similar type of nonreciprocity. However, it has an important advantage compared to metamaterial since antiferromagnetic magnetization pattern is not assumed {\it a priori}, but rather explained by the exchange interaction. On the other hand, our system is conceptually similar to more complex condensed matter systems such as Cr$_2$O$_3$ or MnBi$_2$Te$_4$ and thus can provide insights into the behavior of an entire class of collinear antiferromagnets. Analyzing our model system, we prove that it does feature dual axion response and validate our conclusions by numerical simulations.

The rest of the Article is organized as follows. In Section~\ref{sec:Model} we introduce our model. We proceed with the theoretical derivation of the dual axion response in Sec.~\ref{sec:Theory} and validate this prediction by numerical simulations in Sec.~\ref{sec:Numerics}.
%
%
Next we assess the distinction between the axion $\chi$ and dual axion $\cd$ responses by simulating excitation of the structure by the external sources. The numerical results in Sec.~\ref{sec:Sources} suggest that our system does feature dual axion physics. Finally, we discuss our results and outline further prospects in Sec.~\ref{sec:Discussion}.




\section{Model and governing equations}\label{sec:Model}



We consider a lattice of spins parallel to the $z$ axis assuming Heisenberg-type antiferromagnetic exchange coupling between the nearest-neighbor spins along the $z$ axis and ferromagnetic exchange coupling in the transverse $x$ and $y$ directions. For clarity, we study a tetragonal lattice with the period $a$ along the $z$ axis and $a_{\bot}$ in the transverse $x$ and $y$ directions. Then the interaction energy of a given spin $\vc{s}_m$ with its neighbors is approximated by
\begin{equation} \label{eq:B and Hconnection}
W_m=J\,\mathbf{s}_{m}\cdot\left(\mathbf{s}_{m+1}+\vc{s}_{m-1}\right)-4J_\bot\,\vc{s}_m^2-Ks_{mz}^2\:,
\end{equation}
where $m$ labels the position of the spin in the $z$ direction, $J$ and $J_\bot$ are positive constants describing antiferromagnetic interaction in $z$ direction and ferromagnetic coupling in the transverse directions, respectively, while $K$ is magnetic anisotropy. We assume here that all spins in a given $z=\text{const}$ plane have the same magnitude $\vc{s}_m$. The above interactions stabilize the spin pattern corresponding to the collinear antiferromagnet.

In our analysis, we are interested in the linear response of this structure to the incident time-varying magnetic field ${\bf H}^{in}$, which triggers the precession of the spins around the $z$ axis captured by the semi-classical Landau-Lifshitz-Gilbert equation
\begin{equation} \label{eq:spin evol nonlin}
\frac{d\vc{s}_{m}}{dt}=-\gamma\,[\vc{s}_{m}\times \vc{H}^{loc}_m]-\alpha_0\,\gamma\,\left[\vc{s}_m\times\frac{d\vc{s}_m}{dt}\right]\:.
\end{equation}
Here, $\gamma=\frac{g\,e}{2m_ec} > 0$ is the gyromagnetic ratio for the atoms forming the spin lattice and $g$ is the g-factor. In case of MnBi$_2$Te$_4$ and Cr$_2$O$_3$, those atoms correspond to Mn and Cr, respectively, while $g\approx 2$ in both cases, matching the electron $g$-factor. The term $\propto\alpha_0$ captures  damping of the spin precession.

The local field $\vc{H}^{loc}_m$ acting on the spin $\vc{s}_m$ in the lattice includes several contributions:
\begin{equation}\label{eq:Htot}
\vc{H}^{loc}_m=\vc{H}_m^{exc}+\vc{H}^{in}(\vc{r}_m)+\vc{H}_m^{lat}+\vc{H}_m^{self}\:.
\end{equation}
The first term arises from the exchange and anisotropy contributions to the system energy and is evaluated as
\begin{align}
& \mathbf{H}_{m}^{exc}= - \frac{\partial W}{\partial \mathbf{m}_m} =  \frac{\partial W}{\gamma\cdot\partial \mathbf{s}_m}\notag\\
& = \frac{J}{\gamma}\cdot(\mathbf{s}_{m+1}+\mathbf{s}_{m-1})-\frac{4J_\bot}{\gamma}\,\vc{s}_m-\frac{2K}{\gamma}\,s_{mz}\,\bas\:,
\end{align}
where we take into account the connection between the spin $\vc{s}$ and the magnetic moment $\vc{m}$ via $\vc{m}=-\gamma\,\vc{s}$. Note that the field produced by exchange and anisotropy mechanisms is nonzero even if the spins are not oscillating.

The second term $\vc{H}^{in}(\vc{r}_m)$ is the field of the external electromagnetic wave with the frequency $\omega$ exciting the structure. If the incident field is weak enough, the response of the material is linear and hence the spins oscillate with the same frequency $\om$, i.e.
\begin{equation}\label{eq:linearspin} \vc{s}_m=\vc{s}_m^{(0)}+\vc{s}_m'\,e^{-i\om t}+\left(\vc{s}_m'\right)^*\,e^{i\om t}\:,   
\end{equation}
where $\vc{s}_m^{(0)}$ captures equilibrium spin distribution for antiferromagnet
\begin{equation}\label{eq:equilibriumspin}
    \vc{s}_m^{(0)}=(-1)^m\,s\,\hbar\,\bas\:,
\end{equation}
where $\bas$ is a unit vector along the $z$ axis and $\vc{s}_m'$ presents complex amplitude of oscillating small corrections to the spin. Here, $s$ is the spin of the atoms forming antiferromagnetic lattice. In case of MnBi$_2$Te$_4$ and Cr$_2$O$_3$ it is equal to $5/2$ and $3/2$, respectively. Note that in the linear approximation the vector $\vc{s}_m'$ lies in $Oxy$ plane and has zero $z$ component due to $\vc{s}_m^2=s^2$.


The oscillating spins, in turn, produce scattered electromagnetic wave acting on selected spin $\vc{s}_m$ and given by the sum of the dipole fields of all other spins in the lattice, $\vc{H}^{lat}_m$. This can be separated into two contributions: $\vc{H}^{lat}_m=\vc{H}_m^{l1}+\vc{H}_{m}^{l2}$. One comes from the spins in the same layer and is approximated by
\begin{equation}
\vc{H}_m^{l1}=-\hat{C}\,\gamma\,\vc{s}_m\:, 
\end{equation}
where $\hat{C}$ is the interaction constant of the array calculated in the Supplementary Materials~\cite{Supplement}, Sec.~I. The tensorial structure of the interaction constant is as follows: $\hat{C}=\text{diag}\left(C_{\bot},C_{\bot},C_{||}\right)$. The second dipole term $\vc{H}_m^{l2}$ captures the radiation from all other layers of the spins. Finally, the radiating spin experiences the radiation reaction field described by $\vc{H}^{self}_m=2iq^3/3\,\vc{m}_n$, where $q=\om/c$ and $\vc{m}_n$ is magnetic dipole moment. 

Due to the structure of Eq.~\eqref{eq:spin evol nonlin}, any contribution to $\vc{H}_m^{loc}$ parallel to the spin $\vc{s}_m$ drops out from the equations of motion. As a result, the transverse exchange coupling $J_{\bot}$ is not manifested, while anisotropy and dipole-dipole interactions contribute jointly to a single constant
\begin{equation}\label{eq:Keff}
K_{eff}=K+\frac{\gamma^2}{2}\,\left(C_{||}-C_{\bot}\right)\:.    
\end{equation}

As we seek the linearized dynamics of the spins [Eq.~\eqref{eq:linearspin}], the local field $\vc{H}_m^{loc}$ includes only static and  monochromatic parts with the frequency $\om$:
\begin{equation}
\vc{H}_m^{loc}=\vc{H}_m^{(0)}+\vc{H}_m'\,e^{-i\om t}+\left(\vc{H}_m'\right)^*\,e^{i\om t}\:,
\end{equation}
which results in the linearized equations of motion
\begin{align}\label{eq:Linearized}
& i\om\,\vc{s}_m'=\gamma\,\left[\vc{s}_m^{(0)}\times\vc{H}_m'\right]+\gamma\,\left[\vc{s}_m'\times\vc{H}_m^{(0)}\right]\notag\\
-& i\om\,\alpha_0\gamma\left[\vc{s}_m^{(0)}\times\vc{s}_m'\right]\:.
\end{align}

In the case of a periodic system, Eqs.~\eqref{eq:Linearized} can be solved analytically defining the bulk properties of the material. If the structure is finite, Eqs.~\eqref{eq:Linearized} for spin oscillation amplitude can be solved numerically. We analyze both cases in Secs.~\ref{sec:Theory} and \ref{sec:Numerics}, respectively.

\section{Derivation of electromagnetic response}\label{sec:Theory}


We start by examining a periodic spin lattice. Combining the linearized equations of motion Eq.~\eqref{eq:Linearized} with the equilibrium antiferromagnetic spin distribution in a periodic lattice Eq.~\eqref{eq:equilibriumspin}, we recover
%
%
\begin{equation} \label{eq:spinmatrix}
\begin{split}
i \omega \mathbf{s}_m' =-(-1)^m\,\alpha\,i\om\,\bas^\times\vc{s}_m'+(-1)^m s\,\hbar\,\vc{e}_z^\times\,\\ \left[J\left(\mathbf{s}_{m-1}' + \mathbf{s}_{m+1}'\right)+2\,(J+K_{eff})\,\vc{s}_m' + \gamma \mathbf{H}_{m} \right]\:. 
\end{split}
\end{equation}
Here, $\alpha=\alpha_0\,\gamma s \hbar$ is the dimensionless damping constant and $\vc{H}_m\equiv\vc{H}^{in}(\vc{r}_m)+\vc{H}_m^{l2}$ is the magnetic field in the structure oscillating as $e^{-i\om t}$ due to the incident wave. The matrix $\vc{e}_z^\times$ is defined as
\begin{equation}
\vc{e}^{\times}_z=
    \begin{pmatrix}
    0&&-1\\1&&0
    \end{pmatrix}\:.
\end{equation} 

Due to the periodicity of the system, the time-dependent perturbations of the spins $\vc{s}_m'$ satisfy Bloch's theorem. Given the positions of the spins $\vc{r}_m=m\,a/2\,\vc{e}_z$, we recover 
\begin{equation} \label{eq:spin amplitudes}
    \vc{s}'_m = \begin{cases}
        \vc{P} e^{i k_z m a / 2}, & m \ \vdots\ 2\:, \\
        \vc{Q} e^{i k_z m a / 2}, & m \not\vdots\  2\:.
    \end{cases}
\end{equation}
Equation~\eqref{eq:spinmatrix} then yields a system of linear equations for $\vc{P}$ and $\vc{Q}$ amplitudes: 
\begin{align}
& \left[-i\om\bas^\times+i\alpha\om-2s\hbar\left( J+K_{eff}\right)\right]\,\vc{P}\notag\\
& -2s\hbar J\,\cos\left(\frac{k_z a}{2}\right)\,\vc{Q}=\gamma\,s\hbar\,\vc{H}_{m(e)}\,e^{-ik_z m a/2}\:,\label{eq:spindyn1}\\
& \left[i\om\bas^\times+i\alpha\om-2s\hbar\,\left(J+K_{eff}\right)\right]\,\vc{Q}\notag\\
& -2s\hbar\,J \cos\left(\frac{k_z a}{2}\right)\,\vc{P}=\gamma\,s\hbar\, \vc{H}_{m(o)}\,e^{-ik_z m a/2}\:,\label{eq:spindyn2}
\end{align}
where $\vc{H}_{m(e)}$ and $\vc{H}_{m(o)}$ stand for the magnetic field at even and odd sites of the spin lattice, respectively.

In turn, magnetic fields $\vc{H}(\vc{r})$ and $\vc{B}(\vc{r})$ also satisfy Bloch's theorem and hence can be expanded as
\begin{equation} \label{eq:field floquet exp}
    \begin{pmatrix} \vc{H}(\vc{r}) \\ \vc{B}(\vc{r}) \end{pmatrix}=\sum_{n}  \begin{pmatrix} \vc{H}_n \\ \vc{B}_n \end{pmatrix} \exp \left[ i(k_z + n b) z \right],
\end{equation}
where $b=2\pi/a$ is reciprocal lattice period. Strictly speaking, the expansion Eq.~\eqref{eq:field floquet exp} should be three-dimensional and take into account the discreteness of the lattice in the transverse $x$ and $y$ directions. However, for conceptual clarity, we simplify the problem and assume uniformly distributed magnetization in $Oxy$ plane. This approximation is justified by the fact that the period $a$ along the $z$ axis is much larger than the transverse period $a_{\bot}$ of the lattice for such materials as MnBi$_2$Te$_4$ or Cr$_2$O$_3$.

In the expansion Eq.~\eqref{eq:field floquet exp} $\vc{H}_0$ and $\vc{B}_0$ harmonics provide the averaged field used for the effective medium description, while $\vc{H}_n$ and $\vc{B}_n$ with $n\not=0$ describe  field components with rapid spatial oscillations. Next we inspect the field $\vc{H}$ at points $\vc{r}_m=ma/2$ with even and odd $m$:
\begin{equation}
\begin{split}
\vc{H}(\vc{r}_{m})=\sum\limits_{n}\,\vc{H}_n\,\exp\left[i(k_z+nb)\,ma/2\right]\\
=\sum\limits_{n}\,\vc{H}_n\,e^{ik_z m a/2}\,e^{i\pi m n}
\end{split}
\end{equation}
and recover the following expressions:
\begin{gather}
\vc{H}_{m(e)}\equiv \left.\vc{H}(\vc{r}_m)\right|_{\text{even}}=\vc{U}\,e^{ik_z m a/2}\:,\\
H_{m(o)}\equiv\left.\vc{H}(\vc{r}_m)\right|_{\text{odd}}=\vc{V}\,e^{ik_z m a/2}\:,
\end{gather}
where $\vc{U}$ and $\vc{V}$ amplitudes are constructed from the Floquet harmonics of magnetic field as
\begin{equation} \label{eq:u v definition}
    \vc{U} = \sum_n \vc{H}_n, \quad \vc{V} = \sum_n (-1)^n\, \vc{H}_n\:.
\end{equation}
On the other hand, the distribution of magnetic field should satisfy the wave equation
\begin{equation}\label{eq:waveeq}
    \nabla \times \nabla \times \vc{H} + \frac{\eps}{c^2}\ds{\vc{B}}{t}= \frac{4 \pi}{c} \nabla \times \vc{j},
\end{equation}
where $\vc{j}$ is the external electric current exciting the structure and having spatial dependence $\vc{j}_0\,e^{ik_z z}$. We also assume isotropic and homogeneous permittivity $\eps$ which  captures qualitatively the effect of the background, e.g. the influence of Bi and Te atoms on the spins of Mn atoms forming an antiferromagnetic lattice in MnBi$_2$Te$_4$.

Using the expansion of magnetic fields $\vc{H}$ and $\vc{B}$ via Eq.~\eqref{eq:field floquet exp}, we obtain the connection between the Floquet harmonics $\vc{H}_n$ and $\vc{B}_n$:
\begin{equation} \label{eq:field floquet wave eq}
    (k_z +  n b)^2\,\left(\bas^\times\right)^2 \, \vc{H}_n + \eps q^2\,\vc{B}_n = 0,
\end{equation}
where $n\not=0$ and $q=\om/c$. By definition, \(\vc{B} = \vc{H} + 4 \pi\, \vc{M} \), and hence \(\vc{B}_n = \vc{H}_n + 4 \pi \vc{M}_n\). Floquet harmonics $\vc{M}_n$ are found integrating the magnetization over the unit cell as:
\begin{equation}
    \vc{M}_n = \frac{1}{a} \int_{-a/4}^{3a/4} e^{-i (k_z + nb) z} \, \vc{M}(z) \, dz.
\end{equation}
The integration interval is a unit cell of the lattice which includes two oppositely oriented spins at positions $z_0=0$ and $z_1=a/2$. Still, the result does not change if the unit cell is shifted keeping the same spins inside.

In our model, time-varying magnetization is supplied by the oscillating spins and hence
\begin{equation} \label{eq:magnetization via spins}
     \vc{M}(\vc{r}) = - \frac{\gamma}{a_\bot^2} \sum_{m} \vc{s}'_{m} \delta(z - m a),
\end{equation}
where $a_\bot$ is the lattice constant in the transverse direction. Using Eq.~\eqref{eq:spin amplitudes} for the spin amplitudes, we recover  \(\vc{M}_n = - (\gamma/V_0) (\vc{P} + (-1)^n \vc{Q})\), where $V_0=a\,a_\bot^2$ is the unit cell volume, and thus
\begin{equation} \label{eq:B and H connection}
    \vc{B}_n = \vc{H}_n - \frac{4 \pi \gamma}{V_0} (\vc{P} + (-1)^n \vc{Q})\:.
\end{equation}
As the spins oscillate in $Oxy$ plane, \( \bas \cdot \vc{M}_n = 0\). Additionally, \( \bas \cdot \vc{B}_n = 0\) from Maxwell's equations, which means that \( \bas \cdot \vc{H}_n = 0\) and Eq.~\eqref{eq:field floquet wave eq} is reduced to 
\begin{equation}\label{eq:floquetconnection}
    (k_z + nb)^2 \vc{H}_n = \eps\,q^2 \vc{B}_n.
\end{equation}
Equations~\eqref{eq:spindyn1},\eqref{eq:spindyn2},\eqref{eq:u v definition},\eqref{eq:B and H connection},\eqref{eq:floquetconnection} fully define the structure of the fields and spin distribution in a periodic antiferromagnetic lattice under study.

We examine the solution to these equations in the effective medium limit when $k_z\ll b$ and $q\ll b$ so that the ratio $\xi=q/b\equiv a/\lambda$ is a small parameter. For example, in Cr$_2$O$_3$ and MnBi$_2$Te$_4$ the $\xi$ parameter at magnon resonance frequency is approximately $7.5\cdot 10^{-7}$ and $1.5\cdot10^{-6}$, respectively. This indicates that the effective medium description is perfectly justified, while dipole-dipole interactions in the lattice can be accurately computed in the static limit. In particular, this means that the radiative losses which scale as $\xi^3$ can be safely ignored.

Similarly to the case of a classical metamaterial composed of magnetized layers~\cite{Shaposhnikov2023,Seidov2025}, we apply the perturbation theory in powers of $\xi$. Combining Eqs.~\eqref{eq:B and H connection}, \eqref{eq:floquetconnection}, we evaluate Floquet harmonics $\vc{H}_n$ with $n\not=0$:
\begin{equation}
    \vc{H}_n = - \frac{4 \pi \gamma\, \eps}{V_0}\, \frac{\xi^2}{n^2}\left[\vc{P} + (-1)^n \vc{Q}\right] +  O\left( \xi^3\right).
\end{equation}
Together with Eq.~\eqref{eq:u v definition} this immediately yields an estimate for $\vc{U}$ and $\vc{V}$ amplitudes
\begin{equation}\label{eq:UVest}
\vc{U}=\vc{H}_0+O(\xi^2)\:,\mspace{6mu} \vc{V}=\vc{H}_0+O(\xi^2)\:.
\end{equation}
Making use of Eqs.~\eqref{eq:spindyn1},\eqref{eq:spindyn2}, we evaluate the amplitudes of spin oscillation:
\begin{align}
(\vc{P},\vc{Q})=\frac{\gamma\,s\hbar}{\Delta} &\,\left[\mp i\om\, \bas^\times- i\alpha\om+2s\hbar\,K_{eff}\right.\notag\\
&\left.+4s\hbar\, J \sin^2\left(\frac{k_z a}{4}\right)  \right]\,\vc{H}_0\:,\label{eq:PQappr}
\end{align}
where upper and lower signs correspond to $\vc{P}$ and $\vc{Q}$ spin amplitudes, respectively, and $\Delta$ is defined as
\begin{align}
& \Delta=\om^2\,(1+\alpha^2)+4i\alpha\om\,s\hbar\,\left(J+K_{eff}\right)\notag\\
&-4s^2\hbar^2\,\left[J^2\,\sin^2\left(\frac{k_z\,a}{2}\right)+2J\,K_{eff}+K_{eff}^2\right]\:.\label{eq:Delta}
\end{align}

Using Eq.~\eqref{eq:B and H connection}, we connect $\vc{B}_0$ and $\vc{H}_0$ via $\vc{B}_0=\vc{H}_0-4\pi\gamma/V_0\,(\vc{P}+\vc{Q})$. In line with the established homogenization theory of solid state~\cite{Agranovich} and photonic~\cite{Silveirinha2007,Alu2011,Gorlach2020} lattices, this yields the permeability of antiferromagnetic structure
\begin{equation}\label{eq:Permeability}
\mu=1-\frac{8\pi \gamma^2\,s\hbar}{V_0\,\Delta}\,\left[-i\alpha\om+2s\hbar\,K_{eff}+4s\hbar\,J\,\sin^2\left(\frac{k_z a}{4}\right)\right]\:.
\end{equation}
The zeros of the denominator $\Delta$ describe the dispersion of magnons~-- collective spin excitations. In the absence of losses Eq.~\eqref{eq:Delta} yields:
\begin{equation}\label{eq:MagnonDisp}
\omega_m(k)=2s\hbar\,\sqrt{K_{eff}^2+2JK_{eff}+J^2\,\sin^2\left(\frac{k_za}{2}\right)}\:.
\end{equation}
As the period of the lattice is orders of magnitude smaller than the wavelength, $k_za\ll 1$ and an incident light excites only $k_z\approx 0$ magnons whose frequency reads
\begin{equation}
\omega_m(0)=2s\hbar\,\sqrt{2JK_{eff}+K_{eff}^2}
\end{equation}
and defines the permeability resonance.


Bulk effective permeability is only a part of the effective description. We are interested in the dual axion response of this antiferromagnetic structure which is manifested as a modification to the boundary conditions. For the material possessing axion response $\chi$ and dual axion response $\cd$, the boundary conditions read~\cite{Seidov2025}:
\begin{gather}
\vc{H}_t^{\text{out}}=\vc{H}_{t}-\chi\,\vc{E}_t\:,\label{eq:HBC}\\
\vc{E}_t^{\text{out}}=\vc{E}_{t}+\cd\,\vc{H}_t\:,\label{eq:EBC}
\end{gather}
where $\vc{E}_t$ and $\vc{H}_t$ are the tangential components of the averaged fields inside the medium, while $\vc{E}_t^{\text{out}}$ and $\vc{H}_t^{\text{out}}$ correspond to the outside medium with no axion or dual axion response. In the other words, emergent axion response causes the discontinuity in $\vc{H}_t$, while emergent dual axion response leads to the jump in $\vc{E}_t$.

To derive the boundary conditions for our system, we consider now a semi-infinite lattice with the first spin located at $z=0$. The first spin of the lattice has less neighbors, which slightly modifies its dynamics. However, we neglect this assuming exactly the same dynamics as in the bulk of the crystal. This assumption is validated below in numerical simulations of the finite lattices.

\begin{figure}[t!]
    \centering
    \includegraphics[width=0.9\linewidth]{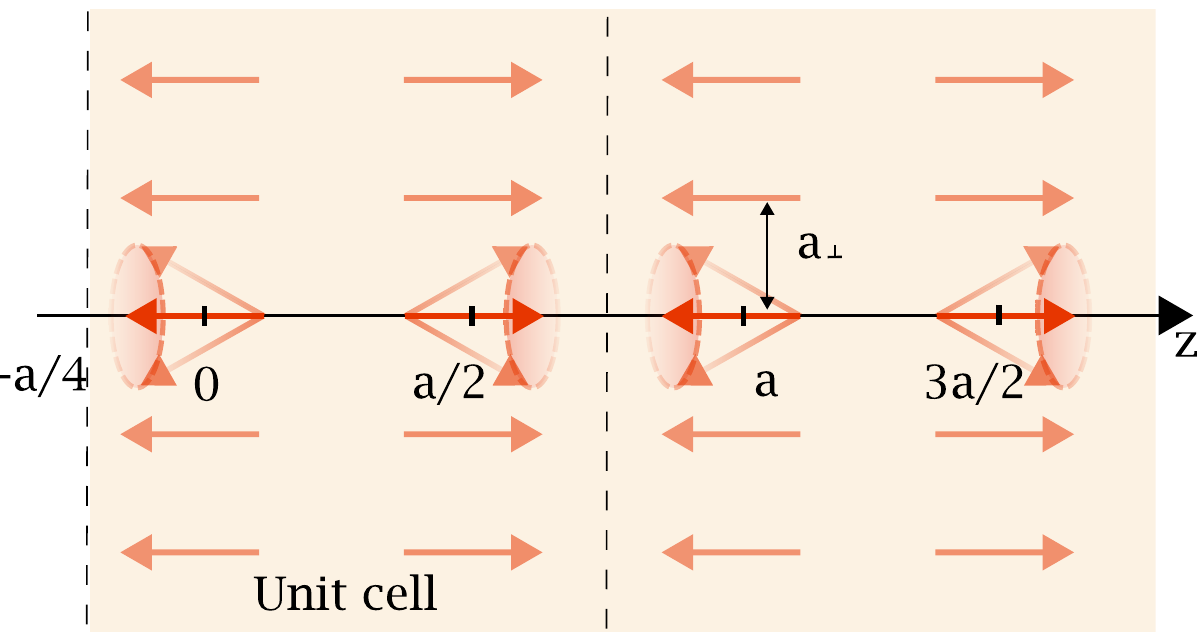}
    \caption{Geometry of antiferromagnetic spin lattice and choice of the material boundary in the effective medium description.} 
    \label{fig:boundaries}
\end{figure}

Replacing this structure by the effective medium [Fig.~\ref{fig:boundaries}], it is crucial to define the position of the boundary. We choose it consistently with the unit cell termination, i.e. at $z_b=-a/4$. The boundary conditions for the microscopic (non-averaged) fields take the usual form
\begin{equation}
    \begin{pmatrix}
        \vc{H}^{\text{out}} \\ \vc{E}^{\text{out}}
    \end{pmatrix} = \sum_{n} \begin{pmatrix}
        \vc{H}_n \\ \vc{E}_n
    \end{pmatrix} e^{i n b z_b}\:,
\end{equation}
where we divided both sides on the same $e^{ik_z z_b}$ phase factor. Similarly to the metamaterial case, we perform the calculation keeping the terms up to the first power in small parameter $\xi$. As Floquet harmonics with nonzero $n$ $\vc{H}_n\sim\xi^2$, $\vc{H}^{\text{out}}(z_b)=\vc{H}_0$ which indicates zero axion response.

At the same time, Floquet harmonics of electric field are non-negligible and are readily evaluated from the Maxwell's equations 
\begin{gather}
    \vc{E}_n = - \frac{1}{q\,\eps}\, (k_z + n b)\, \hat{\vc{e}}_z^\times \vc{H}_n  \notag\\ 
    =\frac{4\pi\gamma}{V_0}\,\frac{\xi}{n}\,\bas^\times\,\left[\vc{P}+e^{i\pi n}\,\vc{Q}\right]\:. \label{eq:Eharmonic}
\end{gather}

Hence, the contribution from the higher-order Floquet harmonics of electric field is non-negligible and has the first order in $\xi$.
\begin{gather}
\sum\limits_n\,\vc{E}_n\,e^{inbz_b}=\sum\limits_n\,\vc{E}_n\,e^{-i\pi n/2}\notag\\
=\vc{E}_0+\frac{4\pi\gamma\xi}{V_0}\bas^\times\left[\vc{P}\sum\limits_{n\not=0}\frac{e^{-i\pi n/2}}{n}+\vc{Q}\sum\limits_{n\not=0}\frac{e^{i\pi n/2}}{n}\right]\:.
\end{gather}
The sums above can be readily evaluated using the Taylor expansion of the logarithm:
\begin{equation*}
    \ln (1-x) = -\sum_{n=1}^\infty \frac{x^n}{n}.
\end{equation*}

This yields
\begin{gather}
\sum\limits_{n\not=0}\,\frac{e^{-i\pi n/2}}{n}=-\sum\limits_{n=1}^{\infty}\,\frac{(e^{i\pi/2})^n}{n}+\sum\limits_{n=1}^{\infty}\,\frac{(e^{-i\pi/2})^n}{n}\notag\\
=\ln (1-i)-\ln(1+i)=\ln(e^{-i\pi/2})=-i\pi/2\:.\notag
\end{gather}
%
%
%
As a result,
\begin{gather}
\sum\limits_{n\not=0}\,\vc{E}_n\,e^{inbz_b}=\frac{2\pi^2\gamma\xi}{V_0}\,i\bas^\times\,(\vc{Q}-\vc{P})\notag\\
=\frac{2\pi\gamma^2\,s\,\hbar\,\omega^2\, a}{c\, V_0\,\Delta}\,\vc{H}_0\:.\notag
\end{gather}
We arrive to the boundary condition of the form $\vc{E}^{\text{out}}=\vc{E}_0+\cd\,\vc{H}_0$ consistent with Eq.~\eqref{eq:EBC} for the dual axion response. We thus conclude that the studied antiferromagnetic spin lattice features the dual axion response 
\begin{equation}\label{eq:DualAxion}
\cd=\frac{2\pi \gamma^2\,s\,\hbar\,\omega^2\,a}{c V_0\,\Delta}\:.    
\end{equation}
The obtained effective material parameters can be additionally simplified by taking into account that $k_za\sim 10^{-6}\ll 1$ and neglecting the spatial dispersion. This yields:
\begin{align}
& \mu=1-\frac{8\pi \gamma^2\,s\hbar}{V_0\,\Delta_0}\,\left[-i\alpha\om+2s\hbar\,K_{eff}\right]\:,\label{eq:Permeabilitysim}\\
& \cd=\frac{2\pi \gamma^2\,s\,\hbar\,\omega^2\,a}{c V_0\,\Delta_0}\:,\label{eq:Cdsim}\\
& \Delta_0=\om^2\,(1+\alpha^2)+4i\alpha\om\,s\hbar\,\left(J+K_{eff}\right)\notag\\
&-4s^2\hbar^2\,\left[2J\,K_{eff}+K_{eff}^2\right]\:.\label{eq:Deltasim}
\end{align}

\begin{figure}
    \centering
    \includegraphics[width=0.85\linewidth]{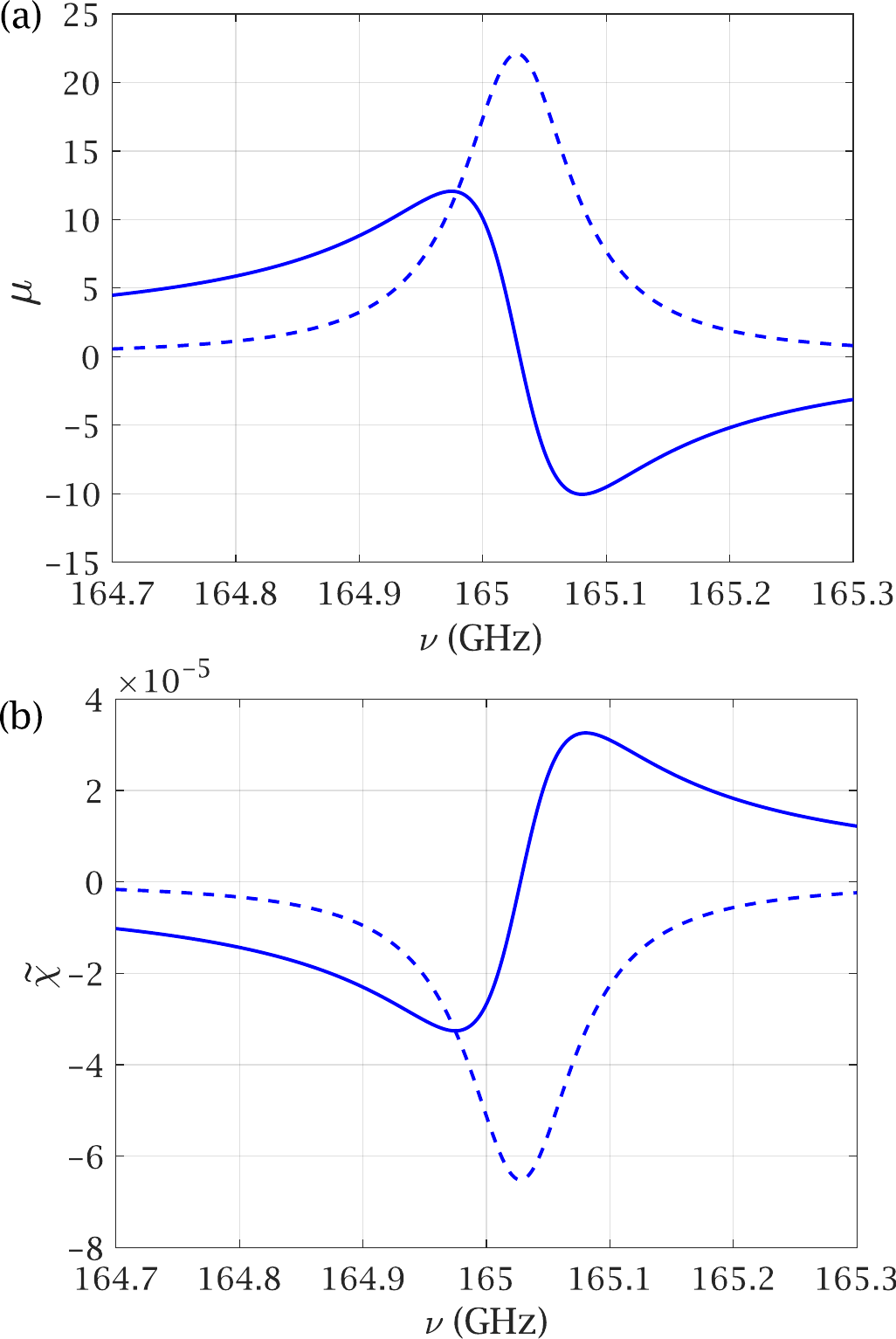}
     \caption{Calculated material parameters of Cr$_2$O$_3$ at frequencies close to magnon resonance. The parameters are $a=1.359$~nm, $a_{\bot}=0.496$~nm, $s=3/2$, $g=2.0$, $\hbar\,K_{eff}/(2\pi)=21.9$~GHz, $\hbar\,J/(2\pi)=58.1$~GHz~\cite{Ovsyannikov2011} and $\alpha=2.2\cdot10^{-4}$~\cite{damping}. Magnon resonance frequency is 165~GHz. (a) Permeability. (b) Dual axion response. Solid and dashed lines indicate real and imaginary parts, respectively.} 
    \label{fig:materialparam}
\end{figure}

To illustrate these results, we consider a celebrated example of material featuring nonreciprocal magneto-electric effect~-- Cr$_2$O$_3$~\cite{Astrov1960,Krichevtsov1993,Kimel2025}. The effective material parameters calculated for this case are depicted in Fig.~\ref{fig:materialparam}. Both permittivity and dual axion response exhibit a resonant enhancement close to magnon resonance at 165~GHz. However, even at resonance, the strength of the dual axion response remains below $4\cdot 10^{-5}$. Note that an analogous estimate for MnBi$_2$Te$_4$~\cite{YanMBT} yields the maximal value of $\cd$ at the same level $\cd\sim 5\cdot 10^{-5}$. Such relatively weak response occurs because the dual axion response is proportional to period-to-wavelength ratio $\xi$, which is below $10^{-6}$ for condensed matter systems. This explains why this effect remains suppressed in condensed matter, while this and analogous phenomena become more pronounced in metamaterials having much larger $\xi$ ratio~\cite{ShuangZhang2025,Gorlach2016}.


\section{Numerical validation}\label{sec:Numerics}

{To validate the derived effective medium description, we examine the scattering of a plane wave at normal incidence from a slab consisting of $N$ pairs of oppositely magnetized layers. We assume that the permittivity of the slab is frequency-independent and constant $\eps=1$, while the permeability $\mu$ and dual axion response $\cd$ are given by Eqs.~\eqref{eq:Permeabilitysim},\eqref{eq:Cdsim}.

The reflection and transmission coefficients are obtained by solving Maxwell’s equations with the modified boundary conditions Eqs.~\eqref{eq:HBC},\eqref{eq:EBC} with $\chi=0$. Defining reflection and transmission matrices via $\vc{E}^{r}=\hat{r}\,\vc{E}^{in}$ and $\vc{E}^t=\hat{t}\,\vc{E}^{in}$, we recover
%
%
\begin{align}
& r_{xx}=r_{yy}=\frac{1}{\Sigma}\,(\tilde{\chi}^2-1+Z^2)\,\sin  \phi\:,\label{eq:rxx}\\ 
& r_{xy}=-r_{yx}=-2\,\tilde{\chi}\,\sin \phi/\Sigma\:,\label{eq:rxy}\\
& t_{xx}=2i\,Z/\Sigma\:, \mspace{6mu} t_{xy}=t_{yx}=0\:,\label{eq:tx}\\
& \Sigma=(\tilde{\chi}^2+1+Z^2) \sin{\phi}+2iZ\cos{\phi}\:.\label{eq:Sigma}
\end{align}
Here \(Z=\sqrt{\mu/\eps}\) is the impedance,  \(\phi=qa\,N\,\sqrt{\eps\mu} \) is optical thickness of the slab in the dimensionless units and $q=\om/c$, while the derivation is presented in Sec.~3 of the Supplementary Materials~\cite{Supplement} and also in Ref.~\cite{Seidov2025}. 

Note that the similar reflection and transmission coefficients introduced for the magnetic field $\vc{H}^{r}=\hat{r}^{H}\,\vc{H}^{in}$ and $\vc{H}^t=\hat{t}^{H}\,\vc{H}^{in}$ are related to the above quantities via
\begin{equation}
\hat{r}^{H}=-\hat{r}\:,\mspace{6mu}\hat{t}^{H}=\hat{t}\:.
\end{equation}

Using Eqs.~\eqref{eq:rxx}-\eqref{eq:Sigma}, we calculate reflection and transmission properties of our model antiferromagnetic structure with the parameters corresponding to Cr$_2$O$_3$ [Fig.~\ref{fig:rteff}]. Characteristic oscillations in the obtained spectra correspond to the standard Fabry-Perot resonances of a finite slab. At frequencies close to magnon resonance the effective permeability increases so that the structure reflects a large part of the incident light. This behavior is manifested as peaks in co-  and cross-polarized reflection accompanied by the minimum in the transmission.

\begin{figure}[ht]
    \centering
    \includegraphics[width=0.85\linewidth]{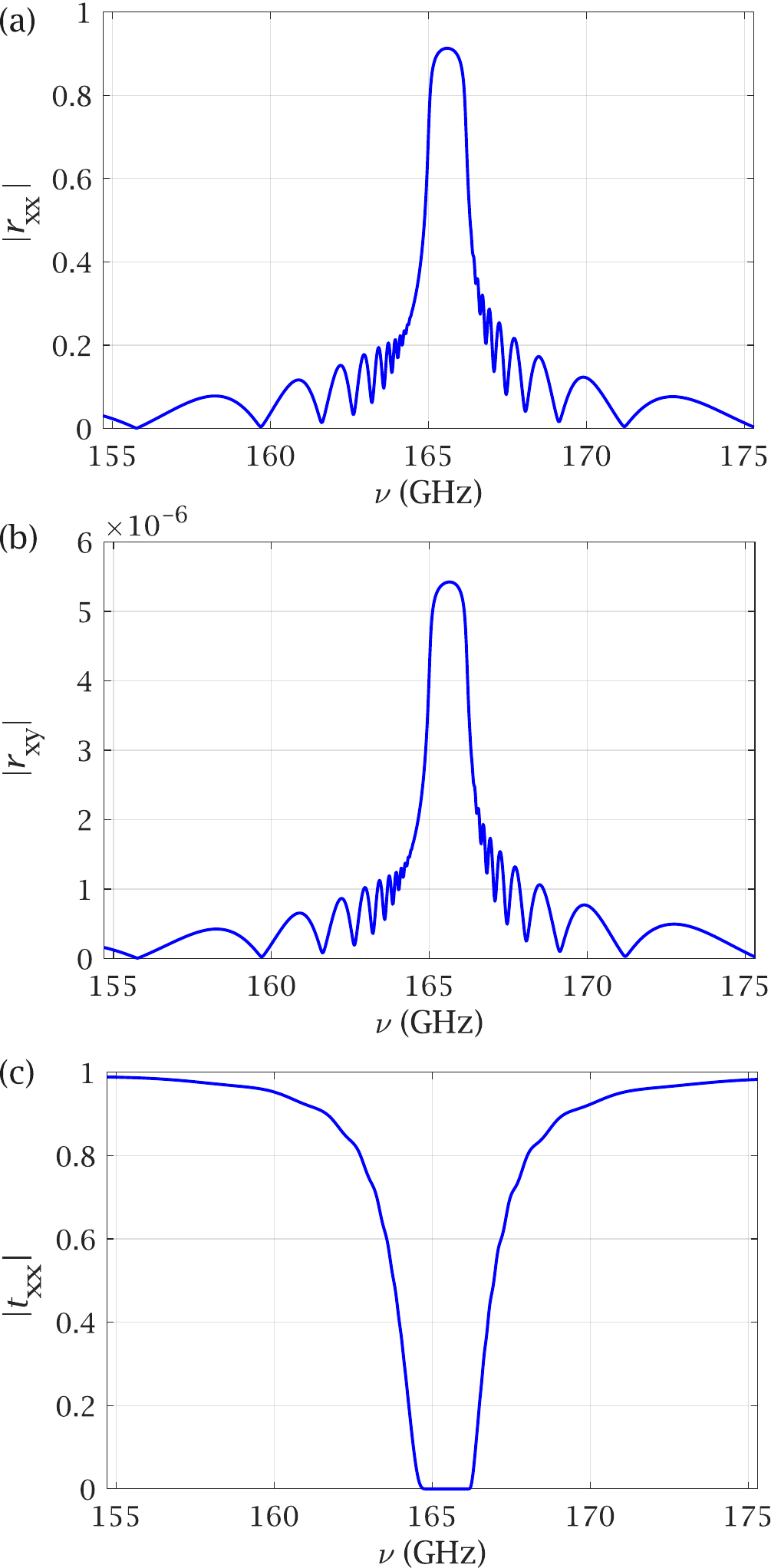}
    \caption{Reflection and transmission properties of a finite slab of model antiferromagnetic structure resembling Cr$_2$O$_3$ calculated in the effective medium approximation. The total number of periods $N=2.0\cdot 10^7$, permittivity $\eps=1$. (a) Co-polarized reflection. (b) Cross-polarized reflection. (c) Co-polarized transmission. Cross-polarized transmission vanishes identically.}
    \label{fig:rteff}
\end{figure}

On the other hand, the same reflection and transmission  coefficients can be computed numerically by solving {the} linearized Landau-Lifshitz-Gilbert equations for the spin dynamics [Eq.~\eqref{eq:spinmatrix}] and evaluating the scattered field produced by the oscillating spins. {This approach corresponds to the discrete dipole approximation widely used in photonics~\cite{Purcell1973,Draine1988,Yurkin2007} and for the analysis of metamaterials in particular~\cite{Shore2007,Belov,Gorlach2014,Chebykin}}. Comparing the results of the effective medium approach to the discrete dipole model, we examine the accuracy of the effective description.

In the framework of the discrete dipole model the local field acting on the $n^{\rm{th}}$ spin is presented as $\vc{H}_n=\vc{H}^{in}(\vc{r}_n)+\vc{H}_n^{lat}+\vc{H}_n^{self}$, while the contributions from exchange interaction and magnetic anisotropy are already taken into account in the equations of motion Eq.~\eqref{eq:spinmatrix}. Here $\vc{H}^{in}$ is the incident field determined by the external sources. In the above scattering problem this is a plane wave. The contributions from the oscillating spins of the lattice and electromagnetic self-action read:
\begin{align}
    &\mathbf{H}_n^{lat}=\frac{1}{a^3}\sum_{n'}\hat{\beta}(n-n')\mathbf{m}_{n'}\:,\\
    &\mathbf{H}_n^{self}=\frac{2iq^3}{3}\mathbf{m}_n\:.
\end{align}
Here $\vc{m}_n=-\gamma\,\vc{s}'_n$ is the magnetic moment associated with the oscillating spin. The self-interaction term $\vc{H}_n^{self}$ describes the radiation reaction field of an oscillating magnetic dipole, see Sec.~4 of the Supplementary Materials~\cite{Supplement}.
The sum with $\hat{\beta}$ captures the field produced by the oscillating rectangular grid of magnetic dipoles in $Oxy$ plane:
\begin{equation}
\hat{\beta}(n-n')=a^3\,\sum_{l,m}\hat{G}(\vc{r}_{l,m,n}-\vc{r}_{l',m',n'})\:,
\end{equation}
$\hat{G}$ being the dyadic Green's function in CGS system of units. Using the dimensionless notations for magnetic field and spin
\begin{equation} 
\mathbf{F} =\frac{\gamma \mathbf{H}}{J\hbar s},\mspace{6mu}  \mathbf{v}_n  = \frac{\mathbf{s}_n}{s\hbar}\:,   \label{notation}
\end{equation}
and introducing the dimensionless parameters
\begin{align}
&\sigma=\frac{2Js\hbar a}{c}\:,\\
& \tau = \frac{2\pi \gamma^2}{J V_0}\:,\\
& \eta = \frac{K}{J}\:,
\end{align}
we derive the following system of linear equations describing spin dynamics:
\begin{align}
&\frac{2iqa}{\sigma}\left[(-1)^{n-1}\bas^\times+\alpha\right]\vc{v}_n = \vc{v}_{n-1} + 2\,(1+\eta)\,\vc{v}_{n} + \vc{v}_{n+1} \notag\\
&+\quad \vc{F}^{in}e^{i k z_n} -\frac{\tau a_{\bot}^2}{2\pi a^2}\,\sum_{n'}\hat{\beta}(n-n')\vc{v}_{n'}
\end{align}
%
%
with $z_n=(n-1)\,a/2$ and $k=q\,\sqrt{\eps}$. The sums $\hat{\beta}(r)$ with nonzero $r$ can be conveniently expressed using Poisson summation in reciprocal space~\cite{Belov,Chebykin} as follows:
\begin{equation}\label{eq:GridField}
\hat{\beta}(r)=2\pi ia\sum_{l,m}[-\vc{k}^\pm_{lm}\otimes\vc{k}^\pm_{lm}+q^2\hat{I}]\frac{e^{\pm ik^z_{lm} ra/2}}{k^z_{lm}}\:.
\end{equation}
The sign $\pm$ corresponds to the sign of  $z$ coordinate of the point where the field is evaluated, $\vc{k}^\pm _{lm}=(k^x_l,k^y_m,\pm k^z_{lm})$.  In a square lattice, $k^x_l=2\pi l/a_\bot$, $k^y_m=2\pi m/a_\bot$, $k^z_{lm}=\sqrt{q^2-(k^x_l)^2-(k^y_m)^2}$. The sign of the square root is chosen such that Im$(k^z_{lm})\geq 0$ which ensures the decay of the evanescent waves with the distance. The series in Eq.~\eqref{eq:GridField} converges exponentially,  which guarantees that only few terms with nonzero $l$ and $m$ are needed for the accurate evaluation of the sum. Essentially, in the deeply subwavelength limit studied here only the terms corresponding to the propagating fields need to be considered.


It is less trivial to compute the sum $\hat{\beta}(0)$ which captures the action of all dipoles in a planar grid on a given dipole in the same lattice. Efficient algorithms to evaluate such sums have been elaborated in the past~\cite{Belov,Chebykin}, while in the quasi-static limit all components of $\hat{\beta}(0)$ are evaluated analytically (see Supplementary Materials, Sec.~I).

The final step is the calculation of reflected and transmitted field produced by the lattice of oscillating spins. This is readily accomplished by summing the contributions from all planes parallel to $Oxy$ and labelled by the index $n$ while keeping in the sums Eq.~\eqref{eq:GridField} only the propagating far-field part. In the dimensionless notations the result reads:
\begin{eqnarray}
    \mathbf{F}^{\text{t}} & = & \mathbf{F}^{\text{in}} - i\tau k_z a\sum_{n=1}^{N-1}  \; \mathbf{v}_n \exp \Big[ i \,k_z\, (z- z_n) \Big]\:, \label{:trdimen}\\
    \mathbf{F}^{\text{r}} & = &-  i\tau k_za\sum_{n=1}^{N-1}  \; \mathbf{v}_n \exp \Big[ -i \,k_z\, (z- z_n) \Big]\:. \label{:refdimen}
\end{eqnarray}
Reflection and transmission matrices are then defined as $\vc{F}^{r}=\hat{r}\,\vc{F}^{in}$, $\vc{F}^{t}=\hat{t}\,\vc{F}^{in}$. Similarly to the effective medium calculation, we obtain $t_{xy}=t_{yx}=0$ for all parameters, which confirms consistency of our numerical procedure.

\begin{figure}
    \centering
    \includegraphics[width=0.85\linewidth]{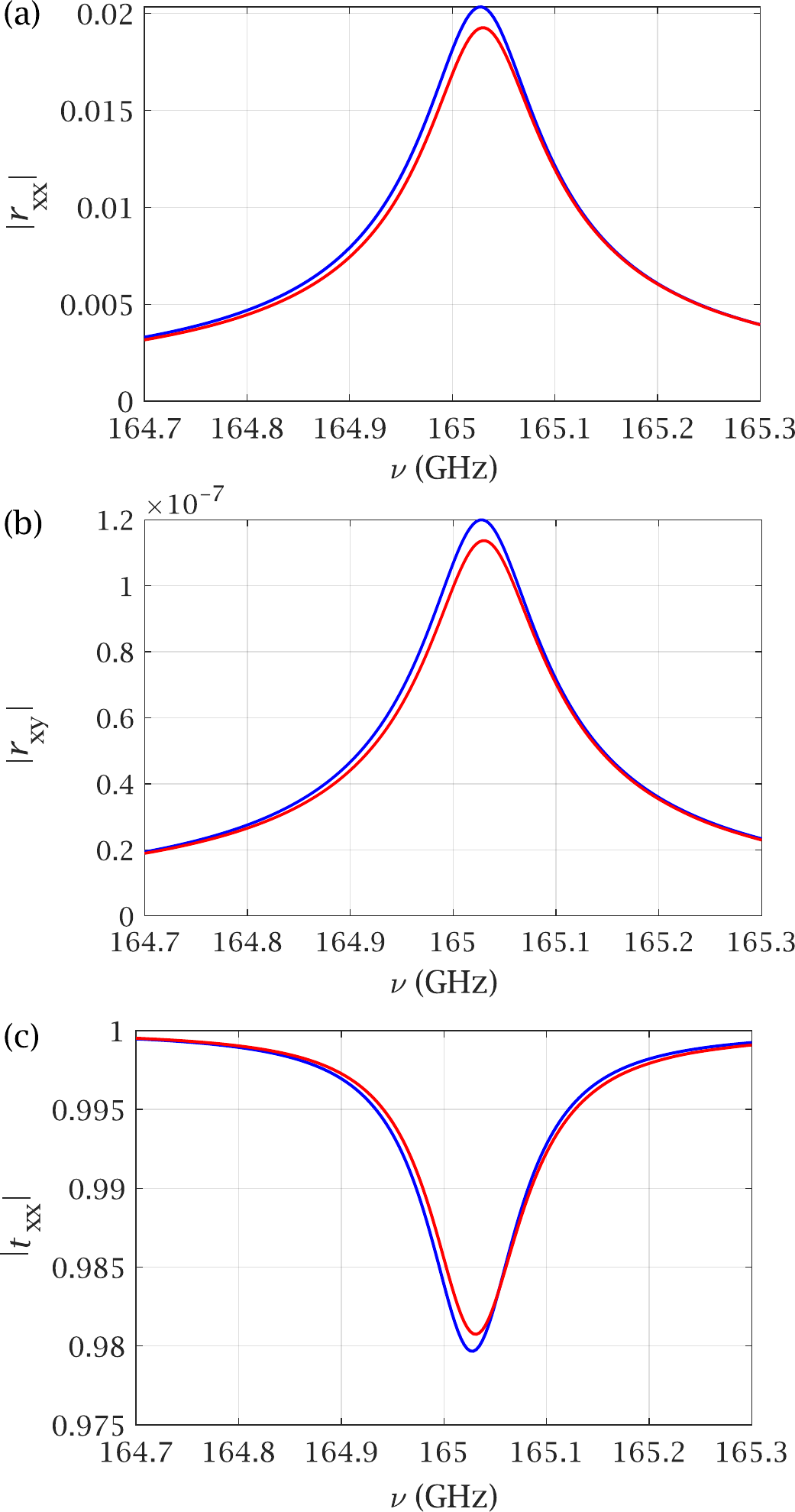}
    \caption{Reflection and transmission properties of a finite slab of model antiferromagnetic structure resembling Cr$_2$O$_3$ calculated via effective medium model (solid blue lines) and discrete dipole method (solid red lines). The total number of periods is $N=800$, while the remaining parameters are the same as in Fig.~\ref{fig:rteff}.}
    \label{fig:refdiscrete}
\end{figure}

To illustrate these results and compare the discrete dipole and effective medium models, in Fig.~\ref{fig:refdiscrete} we plot reflection and transmission coefficients for the plane wave at normal incidence. We compute the scattering coefficients using effective medium approach [Eqs.~\eqref{eq:rxx}-\eqref{eq:Sigma}] with the material parameters Eqs.~\eqref{eq:Permeabilitysim}-\eqref{eq:Cdsim}, see solid blue lines in Fig.~\ref{fig:refdiscrete}. These results are compared to the predictions of the discrete dipole model depicted in Fig.~\ref{fig:refdiscrete} by red solid lines. We observe that the two models agree with a great precision, which is due to the deeply subwavelength nature of the studied material.

While the parameters of the structure correspond to those in Fig.~\ref{fig:rteff}, the overall thickness is taken  much smaller to expedite the discrete dipole calculation. For such thickness, the reflection is much weaker, while the transmission is much closer to unity in the entire studied frequency range. In addition, Fabry-Perot resonances are no longer visible because of the small thickness of the material.



\section{Probing the response by external sources}\label{sec:Sources}

While non-zero cross-polarized reflection [Fig.~\ref{fig:rteff}(b), \ref{fig:refdiscrete}(b)] and zero cross-polarized transmission indicate nonreciprocal magneto-electric effect, these results alone do not allow one to distinguish whether the material features axion or dual axion physics~\cite{Seidov2025}. To identify the type of response in our case and confirm our theoretical prediction independently, we examine the excitation of the spin lattice by the external current inserted directly inside the structure.

The simplest example of such kind is a planar source with the current distribution of the form ${\bf j}=j_e\,\delta(z)\,{\bf e}_{x}$. Placed in vacuum, such source produces a plane wave with the fields
\begin{align*}
& E_x=-\frac{2\pi  j_e}{c}e^{iq|z|}\:,\\
& H_y=-\frac{2\pi j_e }{c}\textnormal{sgn} {(z)}e^{iq|z|}\:,
\end{align*}
i.e. the electric field of the radiated wave is parallel to the oscillating current.

When placed in the middle of the slab of isotropic medium with the admittance $Y=\sqrt{\eps/\mu}$ and an axion response $\chi$, such source also produces a plane wave. However, the polarization of the wave outside of the slab is rotated with respect to the current ${\bf j}=j_e\,{\bf e}_x$ (see Supplementary Materials, Sec.~6) as captured by the expressions 
\begin{align}
E_x=-\frac{4 \pi j_e}{c} \frac{i\, Y \cos{(\phi/2)}+\sin{(\phi/2)}}{2iY\cos \phi + (1+Y^2+\chi^2)\sin \phi}\:,\label{eq:AxionSource1}\\
E_y=-\frac{4 \pi j_e}{c} \frac{\chi\,\sin(\phi/2)}{2iY\cos \phi + (1+Y^2+\chi^2)\sin \phi}\:,\label{eq:AxionSource2}
\end{align}
where $\phi=qa\,N\,\sqrt{\eps\,\mu}$ is optical thickness of the slab and $N$ is the number of periods (i.e. $Na$ is the total thickness). The ratio between the components $E_y$ and $E_x$ captures the strength of polarization rotation and is proportional to the axion response $\chi$.

A similar effect happens in the slab of the material with the dual axion $\cd$ response. Straightforward calculation ({See Supplementary Materials~\cite{Supplement}, Sec.~6}) yields:
\begin{align}
& E_x=-\frac{4 \pi j_e}{c} \frac{i Y \cos{(\phi/2)}+\sin{(\phi/2)}}{2iY\cos \phi  + (1+Y^2+\tilde{\chi}^2Y^2)\sin \phi}\:,\label{eq:DualSource1}\\
& E_y=-\frac{4 \pi j_ e}{c} \frac{i\tilde{\chi}\, Y\cos {(\phi/2)}}{2iY\cos{\phi} + (1+Y^2+\tilde{\chi}^2Y^2)\sin \phi}\:,\label{eq:DualSource2}
\end{align}
with the same identification of $\phi$ and $Y$ parameters.

\begin{figure}
    \centering
    \includegraphics[width=0.85\linewidth]{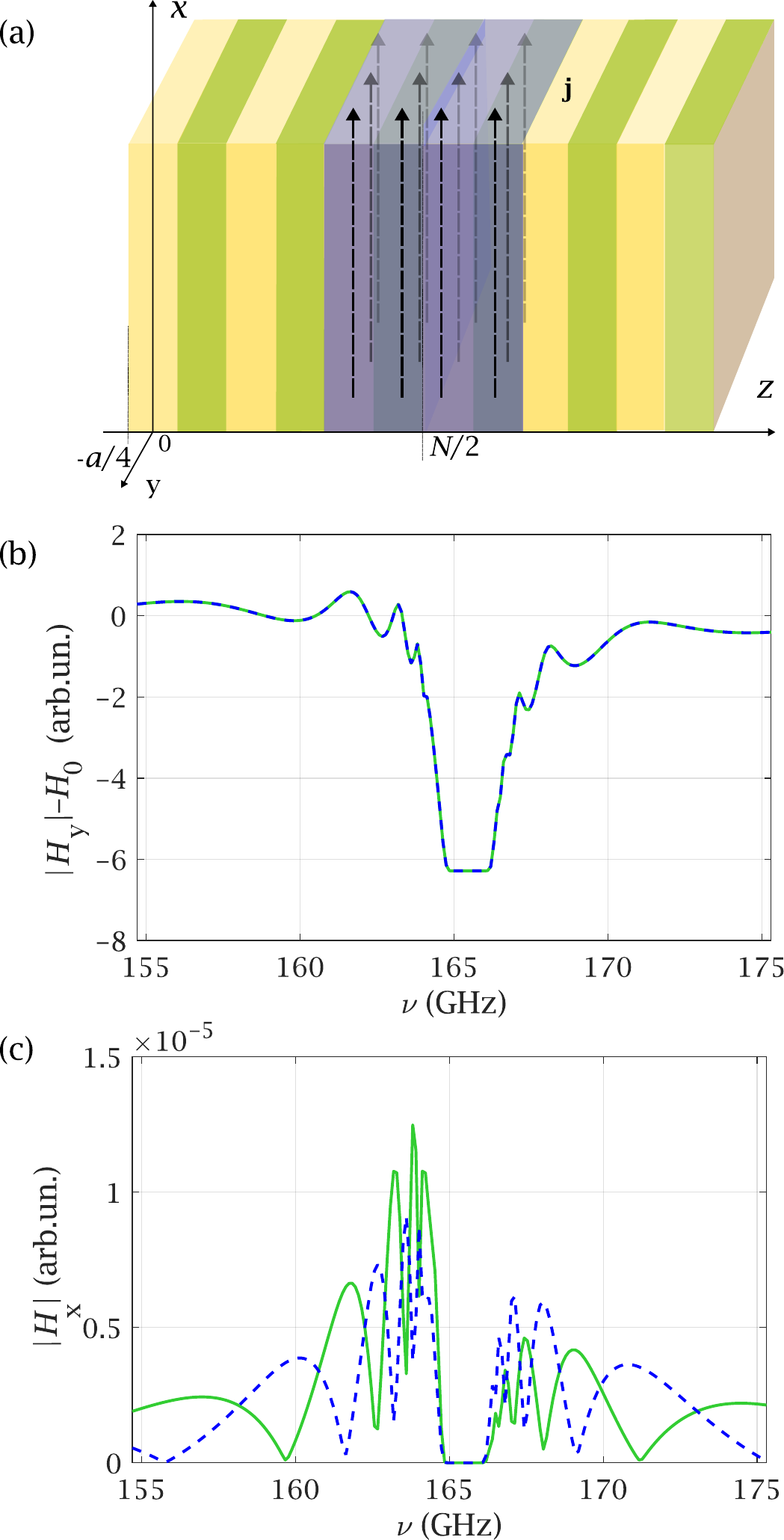}
    \caption{Probing the type of nonreciprocal magneto-electric effect by the external sources. (a) Geometry of excitation of antiferromagnetic spin lattice. (b,c) Co- and cross-polarized components of magnetic field produced by the planar current source. The magnetic field is presented in the arbitrary units. $H_0$ denotes the  magnetic field of a planar current in vacuum.  Green and blue dashed lines correspond to the effective medium calculation for axion and dual axion responses, respectively. Parameters of the simulation match those in Fig.~\ref{fig:rteff}.}
    \label{fig:sourceeff}
\end{figure}

Inspecting Eqs.~\eqref{eq:AxionSource1}-\eqref{eq:DualSource2}, we recover that the frequency dependence of the cross-polarized component $E_y$ produced by the current source is profoundly different in the axion and dual axion cases. In the former situation it is governed by $\sin(\phi/2)$, while in the latter it is proportional to $\cos(\phi/2)$. This distinctive dependence provides us a recipe to identify the type of nonreciprocal magneto-electric coupling in our structure by simulating its excitation by the external sources. Note that instead of electric field one may examine the magnetic field because of the straightforward connection between the two in vacuum.


First we examine again a relatively thick structure of $N=2\cdot 10^7$ periods calculating its excitation by the planar current in the effective medium limit [Fig.~\ref{fig:sourceeff}]. Co- and cross-polarized signals are evaluated for the two scenarios: (a) the structure possesses effective dual axion response given by Eq.~\eqref{eq:Cdsim}, see blue dashed lines; (b) the material features effective axion response $\chi$ producing exactly the same cross-polarized reflection as the dual axion response Eq.~\eqref{eq:Cdsim}, see solid green lines.

The co-polarized field components are the same for both scenarios. However, there is a profound difference in the cross-polarized signals. We observe a clear displacement of Fabry-Perot resonances for axion and dual axion scenarios exactly by half-period, which provides a clear recipe to distinguish the two versions of nonreciprocal magneto-electric response. At the same time, both co- and cross-polarized signals are suppressed at magnon resonance which happens due to losses.

\begin{figure}
    \centering
    \includegraphics[width=0.85\linewidth]{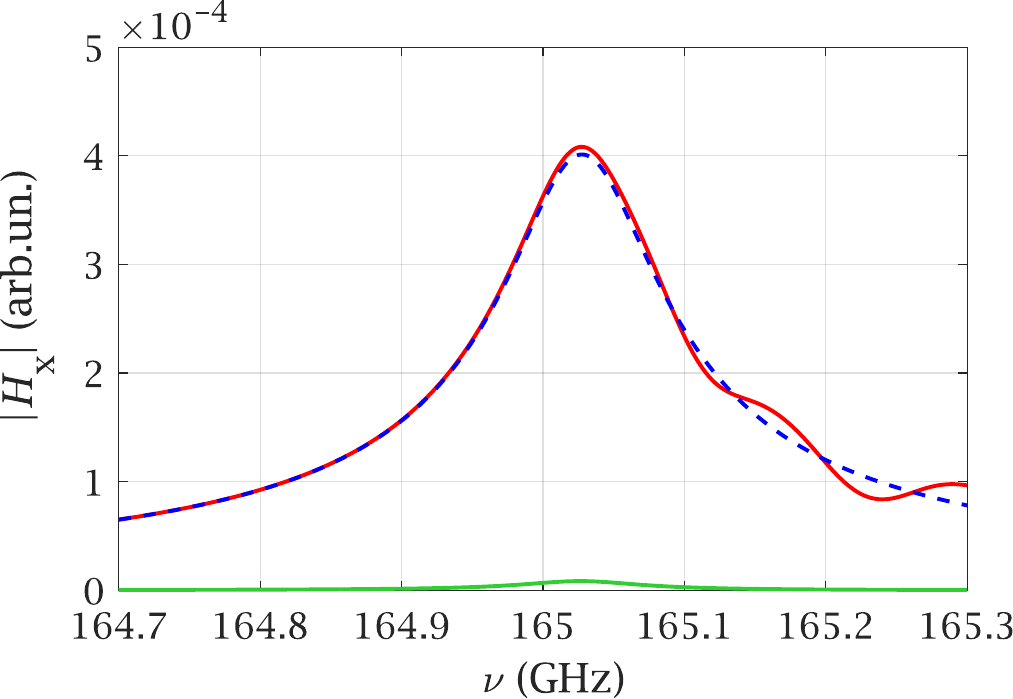}
    \caption{Cross-polarized component of magnetic field created by the current exciting our model antiferromagnetic structure consisting of $N=800$ periods. Solid red line shows the result of the discrete dipole model. Predictions of the effective medium model for the materials with axion or dual axion response are shown by the blue dashed and green solid lines, respectively. Parameters of the simulation match those in Fig.~\ref{fig:refdiscrete}, magnetic field is shown in the arbitrary units having the same scale in all three cases.}
    \label{fig:sourcediscrete}
\end{figure}

To determine which of the two nonreciprocal magneto-electric responses appears in our  structure, we employ the discrete dipole model. Differently from the Sec.~\ref{sec:Numerics}, the external field exciting the structure and causing the precession of the spins is not an incident plane wave, but rather a magnetic field produced by the current source. In the same spirit, we solve the linearized equations of motion for the spins and compute the electromagnetic field they produce. To properly match the effective medium limit, we assume that the current source has a finite thickness $d=2\,a$ [Fig.~\ref{fig:sourceeff}(a)], which allows us to suppress spatial dispersion effects associated with the high-$k$ modes. The implementation of such a finite-thickness current source within the discrete dipole model is discussed in Sec.~7 of the Supplementary Materials~\cite{Supplement}. 

To simplify the discrete dipole calculation, we consider a relatively thin structure consisting of only $N=800$ periods. While Fabry-Perot resonances are not visible any more because of the small slab thickness (Fig.~\ref{fig:sourcediscrete}), the quantitative difference between the axion and dual axion scenarios is well pronounced, see solid green and blue dashed lines in Fig.~\ref{fig:sourcediscrete}. Comparing the two to the discrete dipole model result (solid red line in Fig.~\ref{fig:sourcediscrete}), we confirm that nonreciprocal magneto-electric effect in the structure comes in the dual axion version.

\section{Discussion and conclusions}\label{sec:Discussion}

In summary, we have derived an effective medium description of a model condensed matter structure whose magnetic properties and nonreciprocal magneto-electric response originate from the spin dynamics, while Heisenberg-type exchange interaction mediates antiferromagnetic spin arrangement. Quite surprisingly, we reveal that the consistent description of this  generic system involves electrodynamics with magnetic charge, while nonreciprocal magneto-electric effect is associated with the recently predicted~\cite{Seidov2025} dual axion response.

Despite the simplified nature of our model, these results clearly indicate that recently introduced dual axion response is quite common in condensed matter, Cr$_2$O$_3$ and MnBi$_2$Te$_4$ being potential candidates. As a consequence, our results call for the revision and extension of existing theory of axion electrodynamics in the solids~\cite{Essin2009,Nenno2020,Ahn2022}, while the dual axion response could provide a novel member in the family of topological invariants and topological phases.

Though the correct quantitative predictions for condensed matter structures require more careful description of the spin dynamics and, potentially, first-principles simulations, our study demonstrates that electrodynamics of antiferromagnets is much richer than was originally expected. Electromagnetic properties of recently discovered altermagnets could be another promising direction.





\section*{Acknowledgments}
Theoretical model for the dual axion response in a periodic lattice was supported by Priority 2030 Federal Academic Leadership Project. Numerical simulations were supported by the Russian Science Foundation, grant No.~26-72-10054. Theoretical study of the structure excitation by the external sources was supported by the Ministry of Science and Higher Education of the Russian Federation (Project No. FSER-2025–0012). E.B.-A. acknowledges partial support by the Foundation for the Advancement of Theoretical Physics and Mathematics ``Basis''.

\nocite{Zucker1975,Hardy1920}

\bibliography{dualelina.bib}

\end{document}


\title{Supplemental Materials: \\ Emergence of dual axion response in condensed matter}

\author{Elina Kokurina}
\affiliation{School of Physics and Engineering, ITMO University, Saint  Petersburg 197101, Russia}
\author{Dmitry Vagin}
\affiliation{School of Physics and Engineering, ITMO University, Saint  Petersburg 197101, Russia}
\author{Eduardo Barredo-Alamilla}
\affiliation{School of Physics and Engineering, ITMO University, Saint  Petersburg 197101, Russia}

\author{Maxim A. Gorlach}
\email{m.gorlach@metalab.ifmo.ru}
\affiliation{School of Physics and Engineering, ITMO University, Saint  Petersburg 197101, Russia}


\maketitle
\widetext

\tableofcontents

\section{Dipole-dipole interactions and lattice interaction constant}\label{sec:effmedcorr}

Studying the properties of our antiferromagnetic spin lattice, we take into account dipole-dipole interactions of the spins with each other. At frequencies close to magnon resonance the period-to-wavelength ratio $\xi=a/\lambda$ is of the order of $10^{-6}$ both for Cr$_2$O$_3$ and MnBi$_2$Te$_4$. As a result, the retardation effects are negligible and the dipole field can be evaluated in the quasi-static approximation. The magnetic field of a single magnetic dipole $\vc{m}$ is approximated by the quasi-static formula
\begin{equation}
\vc{H}=\frac{1}{r^3}\,\left[3(\vc{n}\cdot\vc{m})\vc{n}-\vc{m}\right]\:,
\end{equation}
where $r$ is the distance to the observation point and $\vc{n}$ is a unit vector from the dipole to the observation point.

The lattice interaction constant $\hat{C}$ introduced in the main text is evaluated by summing the dipole fields of all spins in the lattice acting on a selected spin in the coordinate origin. Practically, however, the period $a$ is much larger than the period in the transverse directions $a_{\bot}$. We therefore make an approximation and incorporate only the dipole-dipole interactions of the spin with the other spins from the same $Oxy$ plane.

Due to the symmetries of the tetragonal lattice the matrix $\hat{C}$ is diagonal and has the following structure:
\begin{equation}
\hat{C}=\begin{pmatrix}
C_{\bot} & 0 & 0\\
0 & C_{\bot} & 0\\
0 & 0 & C_{||}
\end{pmatrix}\:.
\end{equation}
The term $C_{\bot}$ captures the coupling between the spins, when all of them are aligned along $y$ or $x$ axis. Accordingly, $C_{||}$ describes the interaction for the spins orthogonal to $Oxy$ plane. A straightforward calculation using the quasi-static dipole fields yields:
\begin{equation*}
C_{\bot}=\frac{1}{a_{\bot}^3}\,\sum\limits_{(m,n)\not=0}\,\frac{2m^2-n^2}{(m^2+n^2)^{5/2}}\:.
\end{equation*}
The sum above is convergent and the indices $m$ and $n$ can be swapped. By symmetrizing the expression, we recover
\begin{equation}
C_{\bot}=\frac{1}{2a_{\bot}^3}\,\sum\limits_{(m,n)\not=0}\,\frac{1}{(m^2+n^2)^{3/2}}\:.
\end{equation}
On the other hand, $C_{||}$ term reads
\begin{equation}
C_{||}=-\frac{1}{a_{\bot}^3}\,\sum\limits_{(m,n)\not=0}\,\frac{1}{(m^2+n^2)^{3/2}}\:.
\end{equation}

To proceed, we exploit a useful identity~\cite{Hardy1920, Zucker1975}
\begin{equation}
\sum\limits_{(m,n)\not=0}\,\frac{1}{(m^2+n^2)^{3/2}}=4\,\zeta(3/2)\,\beta_d(3/2)\approx 8\pi \nu_0\:,
\end{equation}
where $\zeta(p)=\sum_{n=1}^{\infty}\,1/n^p$ is Riemann zeta function, \(\beta_d(s)=\sum_{n=0}^\infty (-1)^n/(2n+1)^s\) is the Dirichlet beta function and $\nu_0\approx 0.3594$. Hence
\begin{equation}
C_{\bot}\approx \frac{4\pi}{a_{\bot}^3}\,\nu_0\:,\mspace{6mu} C_{||}=-\frac{8\pi}{a_{\bot}^3}\,\nu_0\:.
\end{equation}

\begin{figure}
    \centering
    \includegraphics[width=0.5\linewidth]{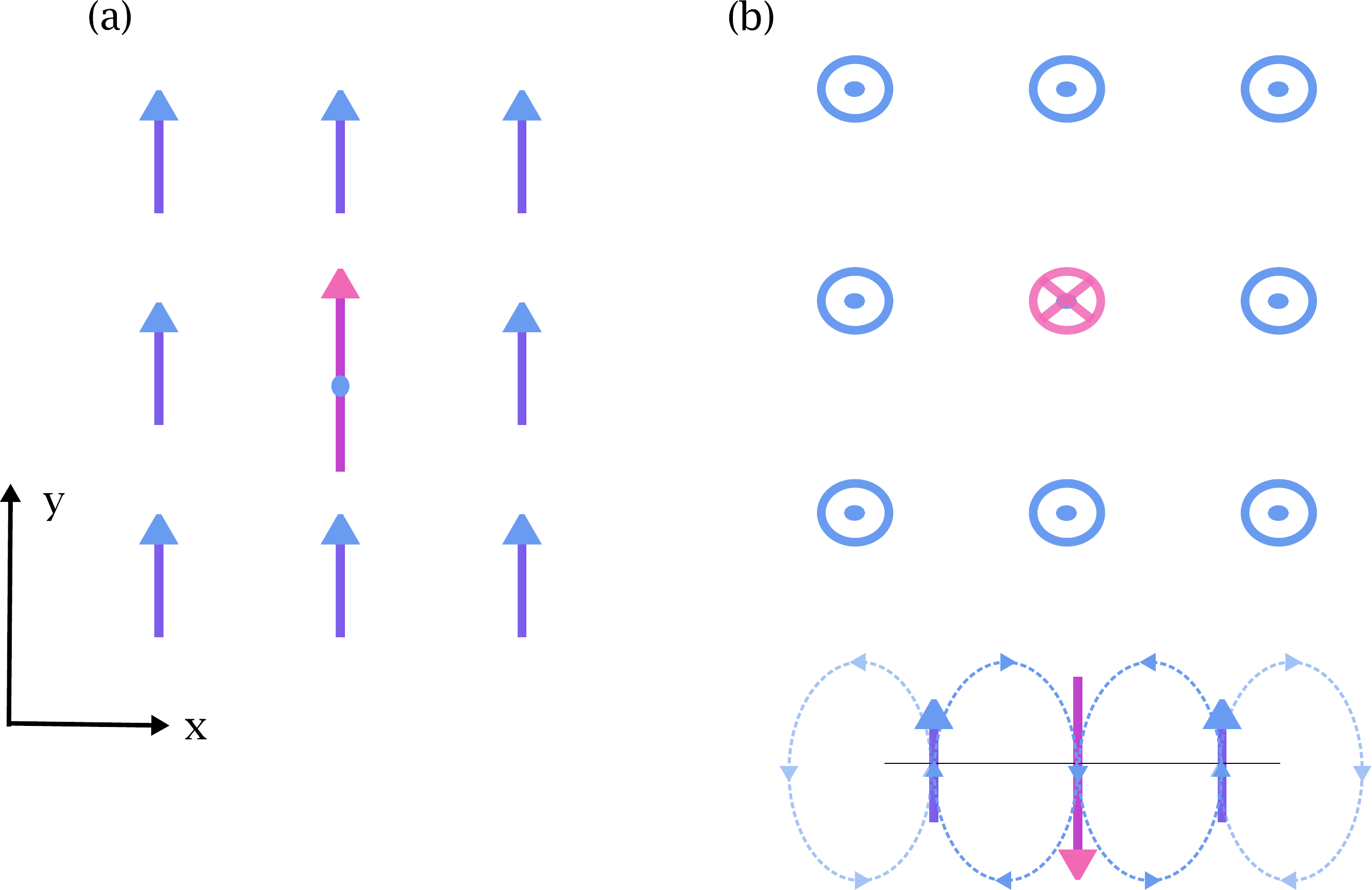}
    \caption{Illustration to the calculation of the lattice interaction constant. (a) Two-dimensional grid of dipoles parallel to the $y$ axis creates the magnetic field aligned in the same direction, i.e. $C_{\perp}>0$. (b) A grid of dipoles orthogonal to the $Oxy$ plane creates the magnetic field antiparallel to the dipoles, i.e.  $C_{\parallel}<0$.}
    \label{fig:placeholder}
\end{figure}

\section{Reflection and transmission properties of axion and dual axion media \label{sec:transref}}

\begin{figure}[b!]
    \centering    \includegraphics[width=0.5\linewidth]{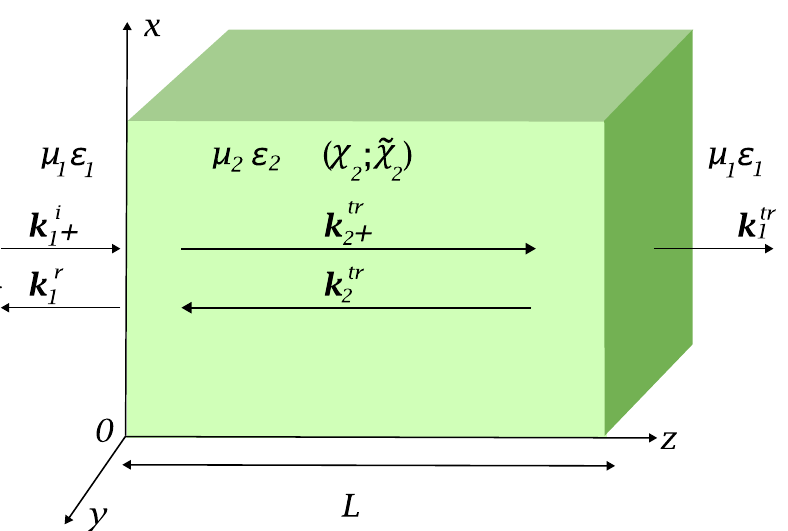}
    \caption{Transmission and reflection of waves through the slab of the continuous medium possessing axion or dual axion response.}
    \label{fig:axion and dual axion}
\end{figure}

In this section, we derive the reflection and transmission coefficients for a slab exhibiting either axion ($\chi$) or dual axion ($\tilde{\chi}$) response within the effective medium approximation. This derivation serves as a consistency check for the results used in the main text and utilizes the transfer matrix technique, see, e.g. Refs.~\cite{Shaposhnikov2023,Seidov2025}. Specifically, we consider a monochromatic plane wave incident normally on a slab of thickness $L$ (Fig.~\ref{fig:axion and dual axion}).

\subsection{Dual axion case $\chi=0,\tilde{\chi}\neq0$}

\begin{figure}[h]
    \centering    \includegraphics[width=0.5\linewidth]{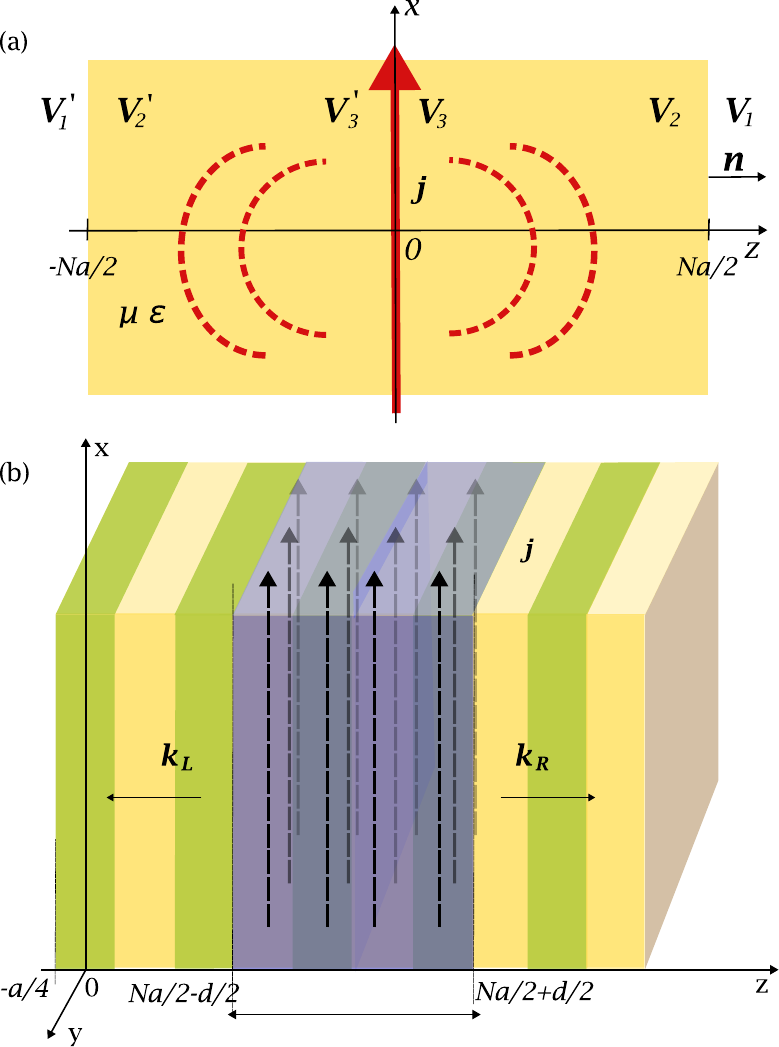}
    \caption{Excitation of the structure by the external sources. 
    (a) Field vectors outside of the medium $\vc{V}_1$, inside the medium close to the boundary $\vc{V}_2$ and close to the external current $\vc{V}_3$. A unit vector $\vc{n}$  is directed in the positive direction of the $z$-axis.
    (b) An excitation scheme by the layer with the distributed currents $\vc{j}$.}
    \label{fig:Discrete currents}
\end{figure}

In a medium with dual axion response, the magnetoelectric coupling modifies the boundary conditions such that the tangential electric field is discontinuous, while the magnetic field remains continuous. For an interface between the two media with parameters $\tilde{\chi}_1$ and $\tilde{\chi}_2$, the boundary conditions read

\begin{align}
\begin{cases}
   &\vc{n}\times(\vc{E}_{1}-\vc{E}_{2})= (\tilde{\chi}_2-\tilde{\chi}_1)\vc{n}\times\vc{H}_2,\\
    &\vc{n}\times(\vc{H}_{1}-\vc{H}_{2})=0.
 \end{cases}
 \label{boundary2}
\end{align}
{Expressing the magnetic field in terms of the electric field via Maxwell equations, the incident $\vc{E}^i$, transmitted $\vc{E}^t$ and reflected  $\vc{E}^r$ waves are related via the matrix equation at the interface:}
\[\begin{pmatrix}
\vc{E}^{t}\\
\sqrt{\frac{\eps_2}{\mu_2}}\vc{E}^{t}
\end{pmatrix}=\hat{M}_{\tilde{\chi}}\begin{pmatrix}\vc{E}^{i}+\vc{E}^{r}\\
\sqrt{\frac{\eps_1}{\mu_1}}\vc{E}^{i}-\sqrt{\frac{\eps_1}{\mu_1}}\vc{E}^{r}\end{pmatrix}
\label{bc}\]
{where boundary transfer matrix is derived from the above boundary conditions:}
\begin{equation}
\label{matdualax}
\hat{M}_{\tilde{\chi}}=
    \begin{pmatrix}
    \hat{I}&&(\tilde{\chi_2}-\tilde{\chi}_1)\hat{\vc{e}_z}^\times\\
    0&&\hat{I}
    \end{pmatrix}
\end{equation}
{Propagation through the slab is described by the bulk transfer matrix coinciding with that in the usual dielectric}
\begin{align}
\hat{M}(z)=\begin{pmatrix} \cos(kz) &-i\sin(kz)\sqrt\frac{\mu}{\eps}&\\-i\sin(kz)\sqrt\frac{\eps}{\mu} &\cos(kz)
\end{pmatrix}; &&\begin{pmatrix}
\vc{E}^{t}(z)\\
\sqrt{\frac{\eps_2}{\mu_2}}\vc{E}^{t}(z)
\end{pmatrix}=\hat{M}(z)\begin{pmatrix}\vc{E}^{i}(0)+\vc{E}^{r}(0)\\
\sqrt{\frac{\eps_1}{\mu_1}}\vc{E}^{i}(0)-\sqrt{\frac{\eps_1}{\mu_1}}\vc{E}^{r}(0)\end{pmatrix}.
\label{bc}
\end{align}
{The full transfer matrix for the dual axion slab of thickness $L$ is obtained by multiplying the respective transfer matrices in the correct order. Subscript $1$ incicates vacuum ($\eps_1=\mu_1=1$, $\cd_1=0$), subscript $2$ corresponds to the material ($\eps_2=\eps$, $\mu_2=\mu$, $\cd_2=\cd$). 
\begin{equation}
\hat{M}=\hat{M}_{\tilde{\chi}-\text{vac}}\hat{M}(L)
\hat{M}_{\text{vac}-\tilde{\chi}}.
\label{finalmatdual}
\end{equation}
{Solving for the reflected and transmitted fields, defined as $\vc{E}^r=\hat{r}\vc{E}^{in}$ and $\vc{E}^{t}=\hat{t}\vc{E}^{i}$, we obtain the reflection and transmission matrices}
%
\begin{align}
& \hat{r} = -(Z_1 \hat{M}_{11}+Z_1 ^2\hat{M}_{12}+\hat{ M}_{21}+Z_1 \hat{M}_{22})^{-1} (Z_1 \hat{M}_{11}-Z_1^2\hat{M}_{12}+ \hat{M}_{21}-Z_1 \hat{M}_{22})\:,\\
& \hat{t}=[2\hat{M}_{11}+Z^2_1\hat{M}_{12}-Z_1\hat{M}_{12}] \hat{R}\:.
\end{align}
where $Z_1 =\sqrt{\mu_1/\eps_1}=1$. The co- and cross-polarized scattering coefficients are 
\begin{align}
r_{xx}&=\frac{(Z^2-1+\tilde{\chi}^2)\sin kL}{\Sigma(\tilde{\chi})}, &
r_{xy}& =-\frac{2\tilde{\chi}\sin kL}{\Sigma(\tilde{\chi})},\\
t_{xx}&=\frac{2iZ}{\Sigma(\tilde{\chi})}, &t_{xy} &=0,
\end{align}
with
\begin{equation}
\Sigma(\tilde{\chi})=2iZ\cos kL+(1+Z^2+\tilde{\chi}^2)\sin kL.
\end{equation}
%
Here, $kL=qL\sqrt{\eps\mu}$ is the optical path inside the slab and $Z=\sqrt\frac{\mu}{\eps}$ is the bulk impedance of the material.

\subsection{Axion case $\chi\neq0,\tilde{\chi}=0$} 

In the axion case, the boundary conditions are modified in the complementary way: the tangential electric field remains continuous, while the magnetic field is discontinuous:
\begin{align}
\label{bcaxion}
\begin{cases}
   &     \vc{n}\times(\vc{E}_1-\vc{E}_2)=0\\
    &    \vc{n}\times(\vc{H}_1-\vc{H}_2)=({\chi}_1-{\chi}_2)\vc{n}\times\vc{E}_1
    \end{cases}
\end{align}
{Following the same procedure, we construct the interface matrix connecting the fields inside and outside of the medium}.
\begin{equation}
\hat{M}_{{\chi}}=
    \begin{pmatrix}
    \hat{I}&&0\\
   ({\chi_1}-{\chi}_2) \hat{\vc{e}_z}^\times&&\hat{I}
    \end{pmatrix}
\label{mataxion}\end{equation}
{The transfer matrix for the entire slab with the axion response reads:}
\begin{equation}
\hat{M}=\hat{M}_{{\chi}-\text{vac}}\hat{M}(L)
\hat{M}_{\text{vac}-{\chi}}
\label{finalmat}
\end{equation}
The reflection and transmission coefficients are
\begin{align}
r_{xx}&=\frac{(-Y^2+1-{\chi}^2)\sin kL}{\Sigma({\chi})}, &
r_{xy}& =\frac{2{\chi}\sin kL}{\Sigma({\chi})},\\
t_{xx}&=\frac{2i\,Y}{\Sigma({\chi})}, &t_{xy} &=0,
\end{align}
with
\begin{equation}
\Sigma({\chi})=2i{Y}\cos kL+({1}+{Y^2}+{\chi}^2)\sin kL,
\end{equation}
where $Y=\sqrt{\eps/\mu}=1/Z$ is the admittance.

\section{Field of a planar current source inside a slab}\label{sec:currenteff}

{As shown in Ref.~\cite{Seidov2025}, axion ($\chi$) and dual axion ($\tilde{\chi}$) responses are indistinguishable in the conventional scattering experiments in the sense that the material possessing permittivity $\eps$, permeability $\mu$ and dual axion response $\cd$ scatters incident electromagnetic waves exactly in the same way as another material with the permittivity $\eps'$, $\mu'$ and axion response $\chi'$. The connection between the two sets of parameters is given by the mapping~\cite{Seidov2025}. As a result, reflection and transmission measurements alone cannot identify the specific type of nonreciprocal magneto-electric coupling.}

{However, the mapping between the axion and dual axion responses breaks down when the source of radiation is placed inside the material, which provides the direct way to distinguish between axion and dual axion physics. To demonstrate this, we consider a planar current source embedded in the middle of the slab and analyze the radiation it produces outside of the slab.}
To describe wave propagation inside the material, it is convenient to introduce circular polarization basis:
\begin{equation}
    \vc{e}_{\pm}={\vc{e}_x \pm i\vc{e}_y}
\end{equation}
and expand the fields as
\begin{align}
& \vc{E}=E_+\vc{e}_{+}+E_{-}\,\vc{e}_{-}\:,\\
& \vc{H}=H_+\vc{e}_{+}+H_{-}\,\vc{e}_{-}\:.
\end{align}
%
The connection to the Cartesian components of the fields is:
\begin{align}
\label{carttocirc}
    &E_x=E_++E_-\:,\\
    &E_y=i(E_+-E_-)\:,
\end{align}
where $E_+$  and $E_-$ correspond to the amplitudes of right- and left-circularly polarized fields, respectively. We introduce a four component field vector
\begin{equation}
\vc{V} = (E_+, H_+, E_-, H_-)^T.
\end{equation}
The key to distinguishing the axion and dual axion responses lies in the boundary conditions. For the sake of generality, we study the material exhibiting both $\chi$ and $\tilde{\chi}$. The  boundary conditions Eqs.~\eqref{matdualax}, \eqref{mataxion} can be  compactly written in the circular basis and in the matrix form
$\hat{M}(\chi, \tilde{\chi})$:
\begin{align}
\hat{M}(\chi_2, \tilde{\chi}_2)
\begin{pmatrix}
&E_{2+}\\
&H_{2+}\\
&E_{2-}\\
&H_{2-}
\end{pmatrix}=\hat{M}(\chi_1,\tilde{\chi}_1)\begin{pmatrix}&E_{1+}\\
&H_{1+}\\
&E_{1-}\\
&H_{1-}\end{pmatrix}, &&\hat{M}(\chi,\tilde{\chi})=
    \begin{pmatrix}
        1&\tilde{\chi}&0&0\\
        -\chi&1&0&0\\
        0&0&1&\tilde{\chi}\\
        0&0&-\chi&1
    \end{pmatrix}.
    \label{currentax}
\end{align}

On the other hand, the fields in the vicinity of the planar current [Fig.~\ref{fig:Discrete currents}(a)] read 
\begin{align}
\begin{cases}
\begin{split}
   &     \vc{E}_3=\vc{E}'_3,\\
    &    \vc{n}\times(\vc{H}_3-\vc{H}'_3)=\frac{4\pi}{c}\vc{j}.
    \end{split}
    \end{cases}
\end{align}
This provides the connection between the vectors in a circular basis:
\begin{equation}
\begin{pmatrix}
&E_{3+}\\
&H_{3+}\\
&E_{3-}\\
&H_{3-}
\end{pmatrix}=\begin{pmatrix}&E'_{3+}\\
&H'_{3+}\\
&E'_{3-}\\
&H'_{3-}\end{pmatrix}+\begin{pmatrix}&0\\
&\frac{4\pi}{c}j^{+}\\
&0\\
&-\frac{4\pi}{c}j^{-}\end{pmatrix}
\end{equation}
which is recast as
\begin{align} \label{V}
    \vc{V_3}=\vc{V_3}'+\frac{4\pi i}{c}\begin{pmatrix}&0\\&j^+\\&0\\&-j^-\end{pmatrix}=&\vc{V_3}'+\frac{2\pi i }{c}j_0\begin{pmatrix}&0\\&1\\&0\\&-1\end{pmatrix}.
\end{align}
Here $\vc{V}_3$ and $\vc{V}'_3$ are the  fields inside the slab from both sides of the current source, $\vc{V}_2$ and $\vc{V}_2'$ are the fields inside the slab close to its boundaries. The connection between $\vc{V}_3$ and $\vc{V}_2$ is given by
\begin{align} \label{V21}
    &\vc{V}_2=\hat{M}_{}(z)\vc{V}_3.
    \end{align}
Finally, $\vc{V}_1$ and $\vc{V'}_1$ are radiated fields measured outside of the slab close to its boundaries [(Fig.\ref{fig:Discrete currents}(a)]. 
\begin{align} \label{V31}
    &\vc{V}_1=\hat{M}(\tilde{\chi},\chi)\vc{V}_2\:,\\
    \label{V32}
    &\vc{V}'_1=\hat{M}(\tilde{\chi},\chi)\hat{M}(-z)\vc{V}'_3\:.
\end{align}
We are interested in $\vc{V}_1$ and $\vc{V}_1'$ vectors which define the fields outside of the slab which can be measured in the experiment.  Combining Eqs.~\eqref{V},\eqref{V21}-\eqref{V32} we obtain the system:
\begin{align}
 \begin{cases}
    &\vc{V}_3-\vc{V}'_3=\frac{2\pi i}{c}j_0 (0,1,0,-1)^T,\\
    &\vc{V}_3=\hat{M}^{-1}(L/2)\hat{M}^{-1}(\tilde{\chi},\chi)(E_+,-iE_+,E_-,iE_-)^T,\\
    &\vc{V}'_3=\hat{M}^{-1}(-L/2)\hat{M}^{-1}(\tilde{\chi},\chi)(E'_+,iE'_+,E'_-,-iE'_-)^T,
  \end{cases}
\end{align}
where $\hat{M}^{-1}(\tilde{\chi},\chi)$ obtained from Eq.~\eqref{currentax} is a boundary matrix, while $L$ is the total thickness of the slab. 

Solving the above equations, we recover the fields radiated by the current source placed in the middle of the slab with effective axion or dual axion response.

{For the axion case ($\chi\neq0$, $\tilde{\chi}=0$), we obtain
\begin{align}
E_x &= -\frac{4 \pi j_e}{c} \frac{i Y \cos(\phi/2) + \sin(\phi/2)}{2iY\cos \phi + (1+Y^2+\chi^2)\sin \phi},\\
E_y &= -\frac{4 \pi j_e}{c} \frac{\chi \sin(\phi/2)}{2iY\cos \phi + (1+Y^2+\chi^2)\sin \phi}.
\end{align}

For the dual axion case ($\chi=0$, $\tilde{\chi}\neq0$),
\begin{align}
E_x &= -\frac{4 \pi j_e}{c} \frac{i Y \cos(\phi/2) + \sin(\phi/2)}{2iY\cos \phi + (1+Y^2+\tilde{\chi}^2 Y^2)\sin \phi},\\
E_y &= -\frac{4 \pi j_e}{c} \frac{i\tilde{\chi} Y \cos(\phi/2)}{2iY\cos \phi + (1+Y^2+\tilde{\chi}^2 Y^2)\sin \phi}.
\end{align}}
For the  axion and dual axion case simultaneously ($\chi\neq0, \tilde{\chi}\neq0$),
\begin{align}
    &E_{x}=-\frac{4\pi j_e}{c}\frac{(1+\chi\tilde{\chi})[iY\cos(\phi/2)+\sin(\phi/2)]}{2iY\,(1+\chi\tilde{\chi})\cos \phi+[1+Y^2+\chi^2+\tilde{\chi}^2Y^2]\sin{\phi}},\\
    &E_{y}=-\frac{4\pi j_e}{c}\frac{(1+\chi\tilde{\chi})[\chi\sin(\phi/2)+i\tilde{\chi}Y\cos(\phi/2)]}{2iY\,(1+\chi\tilde{\chi})\cos \phi+[1+Y^2+\chi^2+\tilde{\chi}^2Y^2]\sin{\phi}}.
\end{align}}
where $\phi = kL$ is the optical thickness of the slab and $Y=\sqrt{\varepsilon/\mu}$ is the admittance of the medium.

Importantly, the fields of a fixed current distribution in an axion and dual axion slabs are not connected to each other by the mapping and this provides a recipe to distinguish the two versions of nonreciprocal magneto-electric response. Such distinction is demonstrated numerically in Sec.~V of the main text.



\section{Radiation from a finite thickness current source in the discrete lattice}\label{sec:currentdisc}

The calculation in the previous section describes the field of an external planar current source in the continous medium with the effective axion or dual axion response. To compare these results to the microscopic discrete model, we consider the excitation of the spin lattice by a current source embedded directly in the structure.

%


While in the effective medium calculation above we assumed infinitely thin planar current, in the discrete model the source should have a finite thickness~-- otherwise the field it produces will be strongly dependent on the positioning within the unit cell.


To ensure consistency with the effective medium description, we model the current as a finite layer of thickness $d=2a$. This choice provides a clear separation of scales: the source remains thin on the macroscopic level ($d \ll L$), while extending over several lattice sites, suppressing the dependence on positioning and allowing for a proper comparison to the effective medium model [Fig.~\ref{fig:Discrete currents}(b)].

Next we calculate the magnetic field component $H_y$ generated by this extended current at the position of a spin located at $z_n$. The problem separates into three spatial regions relative to the current layer boundaries, 
\[
z_L = \frac{Na}{2} - \frac{d}{2}, \quad
z_R = \frac{Na}{2} + \frac{d}{2}.
\]


%

Region I:   $z_n>z_R $
\begin{align}
    H_I(z_n)=-\int_{z_L}^{z_R}\frac{2\pi j}{cd} e^{iq(z_n-z)}dz=-\frac{4 \pi j}{cqd} e^{iq(z_n-N\frac{a}{2})}\sin\left(q\frac{d}{2}\right)\:.
\end{align}

Region II:  $z_n<z_L $
\begin{align}
    H_{II}(z_n)=\int_{z_L}^{z_R}\frac{2\pi j}{cd} e^{-iq(z_n-z)}dz=\frac{4 \pi j}{cqd} e^{-iq(z_n-N\frac{a}{2})}\sin\left(q\frac{d}{2}\right)\:.
\end{align}

Region III:  $z_L<z_n<z_R$
    \begin{equation}
    H_{III}(z_n)=\frac{4\pi j}{cqd}e^{iq\frac{d}{2}}\sin\left(N\frac{a}{2}-z_n\right)\:.
\end{equation}

This field is used in the linearized Landau-Lifshitz-Gilbert equations as an external field exciting spin precession. The results of the calculation are presented in Sec.~V of the main text.

\bibliography{dualelina}